\documentclass[a4paper,10pt]{article}
\usepackage[dvipsnames]{xcolor}
\usepackage{graphicx,color}
\usepackage{jheppub} 

\usepackage{slashed}
\usepackage[utf8]{inputenc}
\usepackage{hyperref}
\usepackage{float}
\usepackage{caption}
\usepackage{subcaption}

\usepackage{epsfig}
\usepackage{amsmath,amssymb}

\usepackage[normalem]{ulem}

\usepackage{comment}

\hypersetup{
	colorlinks=true,
	citecolor=black,
	filecolor=black,
	linkcolor=blue,
	urlcolor=blue
}

\newcommand{\vol}{\mathcal{V}}
\newcommand{\numod}{h^{1,1}}

\title{Coexisting Flux String Vacua from Numerical\\  Kähler Moduli Stabilisation}

\author[a]{\small Shehu~AbdusSalam,}
\author[b]{\small Christopher~Hughes,}
\author[b,c]{\small Fernando~Quevedo,}
\author[d,e]{\small Andreas~Schachner}

\affiliation[a]{\footnotesize Department of Physics, Shahid Beheshti University, Tehran, Iran}

\affiliation[b]{\footnotesize DAMTP,University of Cambridge, Wilberforce Road,  Cambridge, CB3 0WA, UK}

\affiliation[c]{\footnotesize New York University, Abu Dhabi, PO Box 129188 Saadiyat Island, Abu Dhabi, UAE.}

\affiliation[d]{\footnotesize ASC for Theoretical Physics, LMU Munich, 80333 Munich, Germany}

\affiliation[e]{\footnotesize Department of Physics, Cornell University, Ithaca, NY 14853, USA}

\emailAdd{abdussalam@sbu.ac.ir}
\emailAdd{ch845@cam.ac.uk}
\emailAdd{fq2054@nyu.edu,fq201@cam.ac.uk}
\emailAdd{a.schachner@lmu.de}

\abstract{

We present a comprehensive study of K\"ahler moduli stabilisation in Type IIB flux compactifications, combining advanced numerical techniques with analytical methods. Our JAX-based computational framework enables efficient scanning of the UV parameter space, while incorporating $\alpha'$ corrections, loop and non-perturbative effects, as well as uplift contributions to the scalar potential. The implementation features rigorous vacuum validation protocols derived from analytic results. We apply our methods to explicit flux compactifications on more than 80{,}000 Calabi-Yau threefolds with $h^{1,1}\leq 6$ Kähler moduli. By systematically scanning over a wide range of values of the flux superpotential $W_0$ and the string coupling $g_s$, we find explicit realisations of every established Kähler moduli stabilisation scenario: for $10^{-15} \leq |W_0| \leq 10^{-2}$ we obtain both KKLT-like and Kähler uplifted vacua, while for the broader range $10^{-1} \leq |W_0| \leq 10^2$ we recover LVS as well as LVS-like hybrid solutions. Notably, we discover significant parameter regions where multiple vacua coexist within a single flux potential, including novel configurations pairing AdS, Minkowski, and dS minima with different volume hierarchies. These findings enable, for the first time, the analysis of vacuum decay processes within fixed flux configurations, complementing the established theory of transitions between distinct flux vacua and decays towards decompactification.

}

\begin{document} 

\maketitle
\flushbottom

\section{Introduction}
\label{sec:Intro}

Despite more than two decades of research on flux compactifications in Type IIB string theory, the space of concrete models with all moduli stabilised remains remarkably small compared to the vast number of potential configurations. Over the years, progress has enabled explicit stabilisation of complex structure moduli and the dilaton from quantised fluxes, while parallel progress has been made in systematic approaches to K\"ahler moduli stabilisation, including perturbative and non-perturbative effects \cite{Giddings:2001yu,Kachru:2003aw,Balasubramanian:2005zx, Conlon:2005ki,Conlon:2006wz} (for general reviews see e.g. \cite{Douglas:2006es,Denef:2008wq,McAllister:2023vgy}, for the state of the art see \cite{McAllister:2024lnt}).

Given that moduli stabilisation represents the primary challenge for connecting string theory to observable physics, it is crucial to both expand the catalogue of explicit flux compactification solutions and extract more general principles. Of particular importance is the task of locating de Sitter solutions in physically viable parameter ranges, which would provide compelling evidence for their existence.

Building on prior analytic and numerical studies such as \cite{Westphal:2006tn, AbdusSalam:2007pm, Cicoli:2012vw, AbdusSalam:2020ywo,Demirtas:2021nlu,Dubey:2023dvu}, we develop a comprehensive framework to address problems within the realm of numerical Kähler moduli stabilisation. To handle the exponential increase in numerical cost as the number of moduli grows, we leverage the JAX ecosystem \cite{jax2018github}, using just-in-time (JIT) compilation, and automatic differentiation to compute the full Type IIB scalar potential efficiently. With this implementation we can flexibly specify general expressions for both the superpotential $W$ and the Kähler potential $K$ as well as include additional potential contributions from uplifting.

Let us note that our analysis currently assumes prior stabilisation of complex structure moduli and the dilaton by fluxes, focusing specifically on the K\"ahler sector. 
In the long run, our methods can be integrated with existing software tools like JAXVacua \cite{Dubey:2023dvu} to stabilise all moduli simultaneously, enabling systematic studies of backreaction and cross-coupling effects within a single, unified computational environment.

In this work, we specifically implement a benchmark scenario that incorporates the full set of effects needed for controlled Kähler moduli stabilisation: the non-perturbative terms in $W$ from Euclidean D3-brane instantons and D7-brane gaugino condensation, the leading $(\alpha')^3$ correction to $K$ computed in \cite{Becker:2002nn}, logarithmic volume redefinitions \cite{Conlon:2009kt,Conlon:2010ji}, and a positive uplift term modelling the energy of an anti-D3-brane as in \cite{Kachru:2003aw}. Alternative uplift mechanisms, such as from T-branes \cite{Cicoli:2015ylx}, or additional perturbative corrections, like the one-loop string corrections derived in \cite{Berg:2005ja}, can be incorporated straightforwardly.

Our pipeline begins by selecting a Calabi-Yau geometry from established databases like \cite{Kreuzer:2000xy}, and computes the total supergravity scalar potential for given UV parameters, such as the flux superpotential $W_0$, and the string coupling constant $g_s$. With our routines, we systematically explore vacuum structures across all Calabi-Yau manifolds with $h^{1,1} \leq 6$ K\"ahler moduli obtained from \cite{Crino:2022zjk}. By scanning over values of $W_0$ and $g_s$, a variety of vacua are found for geometries with $h^{1,1} \leq 6$. For small values $|W_0| \lesssim 10^{-2}$, we reproduce supersymmetric AdS KKLT minima \cite{Kachru:2003aw} and non-supersymmetric Kähler uplifted solutions \cite{Balasubramanian:2004uy,Westphal:2006tn}, both of which can be uplifted to de Sitter by standard contributions such as anti-D3-branes. For generic values $|W_0| \gtrsim 10^{-1}$, we recover the familiar LVS-type non-supersymmetric AdS solutions \cite{Balasubramanian:2005zx,Conlon:2005ki}. We further identify hybrid solutions resembling those in \cite{AbdusSalam:2020ywo}, blending traits of different established vacua constructions.

By varying $|W_0|$ continuously, we demonstrate how LVS minima smoothly transition into either KKLT or Kähler uplifted solutions. Strikingly, over a range of intermediate values of $|W_0|$, two distinct local minima coexist in the same scalar potential.\footnote{Recently, coexisting $F$-flat flux vacua for complex structure moduli for the same choice of $3$-form fluxes have been discovered in \cite{Chauhan:2025rdj} via the numerical techniques of \cite{Dubey:2023dvu}.} Upon adding an explicit uplift term to the potential, we find all combinations of AdS, Minkowski, and dS vacua for both branches. This simultaneous realisation of multiple vacua in a single potential, rather than across distinct flux choices as studied e.g. in \cite{Bousso:2000xa}, provides a novel setting for studying vacuum transitions in string compactifications. Depending on parameters, the small-volume minimum can have either higher or lower vacuum energy, enabling cosmological applications such as inflation.

We organise the paper as follows. After a brief introduction in \S\ref{sec:EFT-and-soln-derivations}, we describe our numerical implementation in \S\ref{sec:numerics}. Subsequently, we present our main findings in \S\ref{sec:results}, starting with reproducing known KKLT, LVS and Kähler uplifted solutions (\S\ref{subsec:known-solns-results}) before describing novel configurations like coexisting minima in \S\ref{subsec:double-mins}. In \S\ref{sec:P11169}, we study the behaviour of our numerical solutions under variations of $|W_0|$ (cf.~\S\ref{subsec:vary-W0-results}) as well as implications for stability and tunnelling rates in the landscape (cf.~\S\ref{subsec:tunnelling}).

\section{The four-dimensional EFT}
\label{sec:EFT-and-soln-derivations}

In this section, we briefly describe the four-dimensional EFT implemented in our numerical framework. Further, we summarise three of the established Kähler moduli stabilisation scenarios which will show up later in our numerical results.

\subsection{$\mathcal{N}=1$ Calabi-Yau flux compactifications}
\label{subsec:notation-and-scalar-pot}

We consider Type IIB superstring theory compactified on a Calabi-Yau threefold $X$. Concretely, we focus on Calabi-Yau hypersurfaces $X$ embedded in toric varieties obtained from triangulations of four-dimensional reflexive polytopes in the Kreuzer-Skarke list \cite{Kreuzer:2000xy}. After orientifolding, the resulting $\mathcal{N}=1$ four-dimensional EFT features chiral multiplets whose scalar components contain $h_+^{1,1}$ K\"ahler moduli $T_i$, $h^{2,1}_-$ complex structure moduli $z^\alpha$, and the axio-dilaton $\tau$. For simplicity, we limit our attention to O3/O7 orientifolds with $h^{1,1}_-=0$ such that $h^{1,1}_+=h^{1,1}$. Such orientifolds can be constructed systematically with the methods of \cite{Moritz:2023jdb}, see also \cite{Jefferson:2022ssj}. Throughout, we work with the database of O3/O7 orientifolds with $h^{1,1}\leq 6$ compiled in \cite{Crino:2022zjk}.

Let us introduce a basis $\{\omega^i\}_{i=1}^{h^{1,1}(X)}$ of $H^4(X,\mathbb{Z})$ together with its dual basis $\{\omega_i\}_{i=1}^{h^{1,1}(X)}$ of $H^2(X,\mathbb{Z})$ so that $\int_X\omega^i\wedge  \omega_j={\delta^i}_j$. In this basis, the triple intersection numbers $\kappa_{ijk}$ of $X$ can be expressed as
\begin{equation}\label{eq:trip_intnums}
    \kappa_{ijk}= \int_X \omega_i\wedge \omega_j\wedge \omega_k\, .
\end{equation}
The Kähler cone $\mathcal{K}_X\subset H^{1,1}(X,\mathbb{R})$ of $X$ is furnished by the choices of closed Kähler forms $J$. We parametrise $\mathcal{K}_X$ by Kähler parameters $\{t^i\}_{i=1}^{h^{1,1}(X)}$ so that the Kähler class $J$ can be written as
\begin{equation}\label{eq:Kahler_form}
    J=\sum_i t^i\,\omega_i\, .
\end{equation}
The classical volume of the Calabi-Yau manifold $X$ in Einstein frame is given by
\begin{equation}\label{eq:cyvolume}
    \mathcal{V}^{(0)} = \frac{1}{6}\int_X J\wedge J\wedge J = \frac{1}{6}\kappa_{ijk}t^it^jt^k\, .
\end{equation}

We write the axio-dilaton as $\tau = C_0 + \mathrm{i}\, s$
where the dilaton $s = \mathrm{e}^{-\phi}$ is related to the inverse string coupling. Further, we introduce the 4-cycle volumes $\tau_i$ (in Einstein frame) and $C_4$-axions $\theta_i$ 
\begin{equation}\label{eq:vol-tau-defs}
     \tau_i =  \dfrac{1}{2}\int_{X}\, J\wedge J\wedge\omega_i=\frac{1}{2}\kappa_{ijk}t^jt^k \; ,\quad\theta_i=\int_{X}\, C_4\wedge\omega_i\, ,
\end{equation}
to define the classical (complexified) Kähler moduli
\begin{equation}\label{eq:moduli-defs}
    T_i = \tau_i + \mathrm{i}\,\theta_i \, .
\end{equation}
Let us note that, in general, the real part of $T_i$ is affected by higher order corrections. That is, in the presence of quantum corrections to the Kähler metric $K_{i\bar{\jmath}}$, the relationship between Kähler parameters $t^i$ and holomorphic coordinates $T_i$ changes as described in e.g.~\cite{Cecotti:1988qn,Grimm:2004uq,Robles-Llana:2006hby,Robles-Llana:2007bbv,Grimm:2007xm,Baume:2019sry,Marchesano:2019ifh}, see \cite{McAllister:2024lnt} for a recent discussion.

The hyperplanes of $\mathcal{K}_X$ give rise to inequalities of the form 
\begin{equation}\label{eq:Hineq}
    H_{Ai}\,t^i\geq 0\; , \quad A\geq h^{1,1}(X)\, .
\end{equation}
For the Calabi-Yau threefolds studied in this work, these constraints become more constraining at large $h^{1,1}(X)$ because $\mathcal{K}_X$ becomes more narrow \cite{Demirtas:2018akl}. On top of that, we typically have to impose even stricter bounds on e.g. divisor volumes $\tau_i$ in order to work in the geometric regime where the $\alpha'$ expansion is under control. This will play a crucial role in our numerical analysis of \S\ref{sec:results}.

The moduli fields $\phi^I=(z^\alpha,\tau,T_i)$ in our effective theory interact through an $F$-term scalar potential $V_F$ which is completely specified by a K\"ahler potential $K$ and superpotential $W$. 
It can be written as
\begin{equation}\label{eq:FtermPotential}
V_F = \mathrm{e}^K \left(K^{I\bar{J}} D_I W D_{\bar{J}}\overline{W} - 3|W|^2\right)\, , 
\end{equation}
where $K^{I\bar{J}}$ is the inverse Kähler metric and $D_IW$ the K\"ahler covariant derivative defined as
\begin{equation}
    D_I W = \partial_I W + K_I W\, .
\end{equation}
In what follows, we assume that all complex structure moduli $z^\alpha$ and the axio-dilaton $\tau$ can be stabilised at high scales by solving $D_\tau W=D_{z^\alpha}W=0$ in the presence of a 3-form flux background \cite{Giddings:2001yu}, see e.g. \cite{Dubey:2023dvu,Ebelt:2023clh,Plauschinn:2023hjw,Chauhan:2025rdj} for progress in numerically finding flux vacua.\footnote{Alternatively, non-supersymmetric flux vacua with $D_\tau W\neq 0,\, D_{z^\alpha}W\neq 0$, but $\partial_\tau V_F=\partial_{z^\alpha}V_F=0$ can also be found via the methods of \cite{Krippendorf:2023idy}.}
After integrating out $z^\alpha$ and $\tau$, we work with an effective scalar potential $V$ for the remaining Kähler moduli given by
\begin{equation}\label{eq:scalar-pot}
V = \mathrm{e}^K \left(K^{i\bar{\jmath}} D_i W D_{\bar{\jmath}}\overline{W} - 3|W|^2\right)\, . 
\end{equation}

To compute the scalar potential $\eqref{eq:scalar-pot}$, we have to specify a Kähler potential $K$ and a superpotential $W$.
We define the Kähler potential as
\begin{equation}\label{eq:Kpot-defs}
K = K_{\text{cs}} - \log (-\mathrm{i}(\tau-\bar{\tau})) - 2\log\left( \mathcal{V} \right)\, ,
\end{equation}
in terms of the corrected Calabi-Yau volume in Einstein frame
\begin{equation}\label{eq:Vcor}
    \mathcal{V}=\mathcal{V}^{(0)} +  \delta \mathcal{V}\, .
\end{equation}
Here, $\mathcal{V}^{(0)}$ is the classical Calabi-Yau volume defined in Eq.~\eqref{eq:cyvolume} and $\delta \mathcal{V}=\delta \mathcal{V}(t^i,s)$ incorporates general perturbative corrections to the K\"ahler potential. We reiterate that, for our purposes, $K_{\text{cs}}$ in \eqref{eq:Kpot-defs} is a constant determined by the stabilisation of the complex structure moduli $z^\alpha$.

The superpotential used in our analysis is given by
\begin{equation}\label{eq:Wpot-defs}
    W = W_0 + \sum_{D} A_D\, \mathrm{e}^{-a_D T_D}\, ,
\end{equation}
where $W_0$ is the VEV of the flux superpotential \cite{Gukov:1999ya} again determined by the stabilisation of complex structure moduli. The second term accounts for non-perturbative contributions from D-brane instantons hosted on some effective divisor $D= c^iD_i$, $c^i\in \mathbb{Z}$, such as Euclidean D3-branes and gaugino condensation on seven-branes \cite{Witten:1996bn}. We write $a_D = 2\pi/c_{D}$ in terms of the dual Coxeter number $c_{D}$ of the corresponding gauge theory living on the worldvolume of the brane configuration wrapping $D$.
In general, the sum in Eq.~\eqref{eq:Wpot-defs} ranges over smooth,\footnote{Under certain conditions, there can also be contributions from singular divisors to the non-perturbative superpotential as studied in \cite{Gendler:2022qof}.} rigid divisors $D$ for which the Pfaffian prefactors are non-vanishing \cite{Witten:1996bn}. In addition, the sum also includes ``rigidified'' divisors, i.e., non-rigid divisors for which certain deformation modes have been lifted through suitable gauge fluxes $\mathcal{F}_D$ \cite{Martucci:2006ij,Bianchi:2011qh,Bianchi:2012pn,Bianchi:2012kt}, see also \cite{Cicoli:2012vw,Louis:2012nb}.

\subsection{Explicit setup and Kähler moduli stabilisation}
\label{subsec:known-solns-derv}

Let us now describe the explicit ansatz that we make for the perturbative corrections $\delta \mathcal{V}$ to $K$ as defined in \eqref{eq:Vcor}. For this ansatz, we compute the $F$-term conditions and the scalar potential for the Kähler moduli analytically which we occasionally use to compare our numerical implementation with. Moreover, we review the established Kähler moduli stabilisation procedures, namely  KKLT \cite{Kachru:2003aw}, LVS \cite{Balasubramanian:2005zx} and Kähler uplifting \cite{Balasubramanian:2004uy,Westphal:2006tn}, for our explicit ansatz for $K$ and $W$.

We will work with the Kähler potential including the leading perturbative corrections corresponding to the tree-level $(\alpha')^3$ correction of BBHL \cite{Antoniadis:1997eg,Becker:2002nn} and a logarithmic volume redefinition following the arguments in \cite{Conlon:2009kt,Conlon:2010ji}.\footnote{It is important to note that, unlike the BBHL correction, these logarithmic corrections are not yet fully established in concrete string constructions. As discussed in \cite{Burgess:2020qsc, Cicoli:2021rub}, corrections to the scalar potential scaling as $\mathcal{V}^{-8/3}$ may compete with or even dominate over the BBHL term. For a proposal that employs a resummation of these corrections to achieve perturbative moduli stabilisation, see \cite{Burgess:2022nbx}.}
Explicitly, we can write for the corrections $\delta\mathcal{V}$ 
\begin{equation}\label{eq:BBHL+log-pert-corrections}
    \delta\mathcal{V} = \delta\mathcal{V}_{\text{BBHL}}+\delta\mathcal{V}_{\text{log}}
\end{equation}
in terms of the BBHL correction
\begin{equation}
    \delta\mathcal{V}_{\text{BBHL}} = \frac{\xi}{2} s^{3/2}= \frac{\hat{\xi}}{2}  \, ,
\quad \text{with} \quad 
\xi = -\frac{\zeta(3) \chi(X)}{2 (2\pi)^3} \, ,
\end{equation}
and the logarithmic correction
\begin{equation}\label{eq:log_corr}
    \delta\mathcal{V}_{\text{log}}= - \alpha (\mathcal{V}^{(0)})^{2/3}\ln(\mathcal{V}^{(0)})\, .
\end{equation}
Similar to \cite{Cicoli:2024bwq}, we assume that $\alpha$ is small and positive. It is important to emphasise that the microscopic origin of corrections of the form \eqref{eq:log_corr} remains an open question; for a detailed discussion, see \cite{Conlon:2010ji}.

For the moment, we remain agnostic about the underlying mechanism generating contributions to the superpotential \eqref{eq:Wpot-defs}. In what follows, we assume for simplicity that our basis of $H_4(X,\mathbb{Z})$ includes $h^{1,1}$ rigid divisors contributing non-trivially to the superpotential \eqref{eq:Wpot-defs}, while neglecting commensurate contributions from additional rigid divisors.
Then, computing the inverse Kähler metric explicitly for the ansatz \eqref{eq:BBHL+log-pert-corrections}, the scalar potential can be written as
\begin{equation}\label{eq:master-eqn}
\begin{split}
 \mathrm{e}^{-K} V 
=
&|W_0|^2 \, \phi_0+ 
2\sum_{i}A_i |W_0|\mathrm{e}^{-a_i \tau_i}\cos(a_i \theta_i + \theta_0) 
\left[\phi_0 + \phi_1^{(i)}\right]
\\
&+  \sum_{i,j}A_i A_j \mathrm{e}^{-a_i \tau_i - a_j \tau} \cos( a_i \theta_i -  a_j \theta_j)
\left[\phi_0 + \phi_1^{(i)} + \phi_1^{(j)} + \phi_2^{(i,j)}\right]\, ,
\end{split}
\end{equation}
where the various terms $\phi_0, \phi_1^{(i)},\phi_2^{(i,j)}$ are collected in App.~\ref{App:Kmet}.
This master equation generalises the expressions of \cite{AbdusSalam:2020ywo,Cicoli:2021tzt} by including additional perturbative corrections, see also \cite{McAllister:2024lnt} for similar expressions for $V$ and its derivatives. We stress that our numerical implementation does not use \eqref{eq:master-eqn}, but rather derives \eqref{eq:scalar-pot} directly.

We note that, to accurately determine the axion VEVs, one would need to know the phases of $A_i$ which we typically set to zero by hand. Since finding the axion VEVs is mostly straight forward, we will refrain from commenting on them in what follows.

Below, we revisit the three established Kähler moduli stabilisation procedures KKLT \cite{Kachru:2003aw}, LVS \cite{Balasubramanian:2005zx} and Kähler uplifting \cite{Balasubramanian:2004uy,Westphal:2006tn} in the presence of the perturbative corrections as defined in Eq.~\eqref{eq:BBHL+log-pert-corrections}. 

\subsubsection*{KKLT scenario}\label{subsubsec:KKLT-derv}

The KKLT construction \cite{Kachru:2003aw} provides an explicit procedure for obtaining supersymmetric AdS vacua through non-perturbative effects in the superpotential, with perturbative corrections to the K\"ahler potential playing a secondary role. The solutions can be expressed succinctly by starting from the $F$-term conditions for all K\"ahler moduli
\begin{equation}\label{eq:Fterm-equal-0}
D_i W = \partial_{T_i}W + (\partial_{T_i}K)\, W = 0 \, .
\end{equation}
Early works, such as \cite{Denef:2004dm,Denef:2005mm,Lust:2006zg}, have already provided evidence for the existence of supersymmetric AdS$_4$ vacua satisfying \eqref{eq:Fterm-equal-0}, though making use of special structures and without hierarchical scale-separation. More recently, fully explicit supersymmetric AdS$_4$ minima have been found in explicit Calabi-Yau orientifold compactifications containing $\mathcal{O}(100)$ moduli fields featuring large scale-separation \cite{Demirtas:2021nlu} and warped throats \cite{McAllister:2024lnt}.
These works have provided the tools to numerically construct supersymmetric AdS$_4$ minima for any $h^{1,1}$, see also \S\ref{sec:SOAnumerics}.

Let us compute \eqref{eq:Fterm-equal-0} for our ansatz for the Kähler potential \eqref{eq:Kpot-defs} and the superpotential \eqref{eq:Wpot-defs} by writing
\begin{equation}\label{eq:FtermsExplicit}
    -a_i A_i\,\mathrm{e}^{-a_i T_i}-\dfrac{t^i}{2\mathcal{V}}(1+\delta\mathcal{V}')\biggl [W_0+\sum_{k}\, A_k\,\mathrm{e}^{-a_k T_k}\biggl ] =0\, ,
\end{equation}
where we introduced the shorthand notation $\delta\mathcal{V}'=\partial_{\mathcal{V}^{(0)}}\delta\mathcal{V}$.

We may write this in the form
\begin{equation}\label{eq:FtermsExplicit6}
    T_{i}=\dfrac{1}{a_i}\ln(W_0^{-1})-\dfrac{1}{a_i}\ln\biggl (\epsilon^{i}\biggl [1+W_0^{-1}\sum_{k}\, A_k\,\mathrm{e}^{-a_k T_k}\biggl ]\biggl )\, ,
\end{equation}
where
\begin{equation}
    \epsilon^{i}= -\dfrac{t^i(1+\delta\mathcal{V}')}{a_i A_i\,2\mathcal{V}}\, .
\end{equation}
Provided $|\epsilon^{i}|\ll 1$, this can be further simplified to read
\begin{equation}\label{eq:trivial-moduli-relation-ii}
    T_{i}=\dfrac{1}{a_i}\ln(W_0^{-1})-\dfrac{1}{a_i}\ln\biggl (\epsilon^{i}\biggl [1+\sum_{k}\, A_k\,\epsilon^k+\sum_{j,k}\, A_j \, A_k\,\epsilon^j\,\epsilon^k+\ldots\biggl ]\biggl )\, ,
\end{equation}
which is equivalent to a similar expression in \S5.1 of \cite{Demirtas:2021nlu}.

As long as $|\epsilon^{i}|\ll 1$ is guaranteed, the size of the 4-cycle volumes $\tau_i$ is mainly controlled by $\ln(W_0^{-1})$. Achieving Einstein frame volumes $\tau_i\gtrsim 1$ requires $|W_0|\ll 1$, which requires sufficient tuning through suitable choices of fluxes.  In practice, solutions of this type can be systematically constructed through perturbatively flat vacua \cite{Demirtas:2019sip} or, alternatively, by fine tuned cancellations in the flux superpotential \cite{Chauhan:2025rdj}.

\subsubsection*{Large Volume Scenario}
\label{subsubsec:LVS-derv}

The Large Volume Scenario (LVS) provides an alternative moduli stabilisation framework in Type IIB compactifications, where K\"ahler moduli are stabilised through a combination of non-perturbative effects in the superpotential and perturbative corrections to the K\"ahler potential. 

To simplify the discussion, let us start with the case of two Kähler moduli by setting $h^{1,1}=2$ where we denote the two 4-cycle volumes by $\tau_b$ and $\tau_s$ with volume form
\begin{equation}
    \mathcal{V}^{(0)} = \dfrac{1}{9\sqrt{2}}( \tau_b^{3/2}-\tau_s^{3/2}) \, .
\end{equation}
Then, the analysis begins with the master equation \eqref{eq:master-eqn}, taking the large volume limit where $\tau_b$ is sent to infinity. 
Keeping only the leading order terms in this limit, we find the scalar potential 
\begin{equation}\label{eq:LVS-leading-potential}
    V \mathrm{e}^{-K} = \tilde{h} |W_0|^2 + 2 a_s \tau_s A_s |W_0|\mathrm{e}^{-a_s \tau_s}\cos(a_s \theta_s + \theta_0)   + 12\sqrt{2}\,\sqrt{\tau_s} a_s^2 A_s^2\,\mathcal{V}^{(0)}\, \mathrm{e}^{-2 a_s \tau_s} \, ,
\end{equation}
where $\theta_0$ is the phase of $W_0$, and $\tilde{h}$ is given by
\begin{equation}
    \tilde{h} = \frac{3}{2\mathcal{V}^{(0)}}\, \Bigl(\delta \mathcal{V} + \vol^{(0)}(3\,\vol^{(0)}\, \delta \mathcal{V}'' - \delta \mathcal{V}')\Bigr)
\end{equation}
in terms of $\delta \mathcal{V}''=\partial_{\vol^{(0)}}^2\delta\vol$.
As this potential depends on a single axion $\theta_s$, minimisation along this direction yields $a_s \theta_s = \pi - \theta_0$. The stabilisation of the Kähler moduli depends critically on the sign of $\tilde{h}$. Provided that $\tilde{h}>0$, the first and third term in \eqref{eq:LVS-leading-potential} are positive and compete with the second term in such a way that a minimum materialises at
\begin{equation}\label{eq:LVS_solution}
    \mathcal{V}^{(0)}\sim |W_0|\,\mathrm{e}^{a_s\tau_s}\; , \quad \tau_s\sim 1/g_s\; , \quad \langle V\rangle \sim -\dfrac{|W_0|^2}{\mathcal{V}^3}\, .
\end{equation}

Given that LVS depends on this delicate balance between non-perturbative and perturbative corrections in \eqref{eq:LVS-leading-potential}, the explicit form of the Kähler potential $K$ in \eqref{eq:Kpot-defs} is imperative. In particular, for the perturbative corrections in Eq.~\eqref{eq:BBHL+log-pert-corrections}, we find
\begin{equation}
    \tilde{h} = \frac{3\hat{\xi}}{4\mathcal{V}^{(0)}}+\dfrac{3\alpha}{(\vol^{(0)})^{2/3}} \, .
\end{equation}
Hence, the first term in \eqref{eq:LVS-leading-potential} receives two competing contributions: the BBHL correction $\frac{3 \hat{\xi}}{4\vol^{(0)}}$ (positive for $h^{2,1}>h^{1,1}$) and the logarithmic correction $\frac{3 \alpha}{(\vol^{(0)})^{2/3}}$ (positive for $\alpha > 0$). Clearly, the latter generically dominates unless $\alpha$ is appreciably small.

For compactifications with $h^{1,1} > 2$, the above process leaves $h^{1,1}-2$ moduli $\tau_j$ ($j \neq s,b$) as flat directions at this order. These can be stabilised through two distinct mechanisms: either by higher-order string corrections to the K\"ahler potential, leading to a hierarchy $\tau_b > \tau_j \gg \tau_s$, or through additional non-perturbative effects, producing $\tau_b \gg \tau_j > \tau_s$ \cite{Bansal:2024uzr}. The specific stabilisation pattern depends on the relative strength of these corrections in a given compactification. Throughout, we only focus on additional non-perturbative contributions to $W$ to stabilise the remaining Kähler moduli.

\subsubsection*{K\"ahler uplifting}\label{subsubsec:Westphal-derv}

The third type of solutions appearing frequently in our numerical analysis below were proposed in \cite{Balasubramanian:2004uy} and subsequently refined in \cite{Westphal:2005yz,Westphal:2006tn}. This class of minima of $V$ uses the interplay of $\alpha'$ effects in the Kähler potential and non-perturbative contributions to the superpotential $W$. Indeed, they emerge for values of $|W_0|$ exceeding a critical threshold $|W_0^{\text{max}}|$, beyond which KKLT solutions cease to exist. Specifically, starting from a solution to $D_iW=0$ for sufficiently small $|W_0|$ and increasing the value of $|W_0|$ iteratively, supersymmetry breaking can occur at some $\mathcal{V}^{(0)} > \delta\vol$ when $|W_0| = |W_0^{\text{max}}|$. Then, depending on the range of $|W_0|\gtrsim|W_0^{\text{max}}|$ in which a minimum persists, these non-supersymmetric Kähler uplifted solutions are either anti-de Sitter or de Sitter. 

Since $\mathcal{V}^{(0)}$ at the minimum decreases with increasing $|W_0|$, these minima are typically plagued by issues related to control over perturbative corrections. Although we will not address these concerns in this work, we point out that our framework described in \S\ref{sec:numerics} can be employed to study such control problems quite systemically and efficiently.

In practice, de Sitter solutions of this form have been previously found in \cite{Rummel:2011cd,Louis:2012nb}. In our investigations, we find a special variant of these solutions due to the specific form of $\delta\vol$ as defined in Eq.~\eqref{eq:BBHL+log-pert-corrections}.

\subsubsection*{Classification of different scenarios}

Beyond the three established scenarios KKLT, LVS and K\"ahler uplift mentioned above, it is conceivable that moduli can be stabilised through unknown mechanisms. For example, there can be \emph{hybrid} vacua \cite{AbdusSalam:2020ywo} that exhibit characteristics of both standard LVS vacua and conventional KKLT models.

More generally, it is therefore useful to classify models by the process in which each K\"ahler modulus is stabilised. For instance, one can imagine the following scenarios:
\begin{itemize}
\item All moduli are stabilised by non-perturbative effects (like in KKLT).
\item Some moduli are stabilised by non-perturbative effects but others by perturbative effects as in LVS, including fibre moduli stabilisation \cite{Cicoli:2008gp}.
\item All moduli are stabilised by perturbative effects as in recent proposals such as \cite{Burgess:2022nbx, Cicoli:2024bwq, Antoniadis:2018hqy, Leontaris:2022rzj}.
\end{itemize}
All of these scenarios can be easily studied within our numerical framework, which will be introduced in the next section.

\section{Numerical Kähler moduli stabilisation framework}
\label{sec:numerics}

\begin{figure}[!t]
    \centering
    \includegraphics[width=\linewidth]{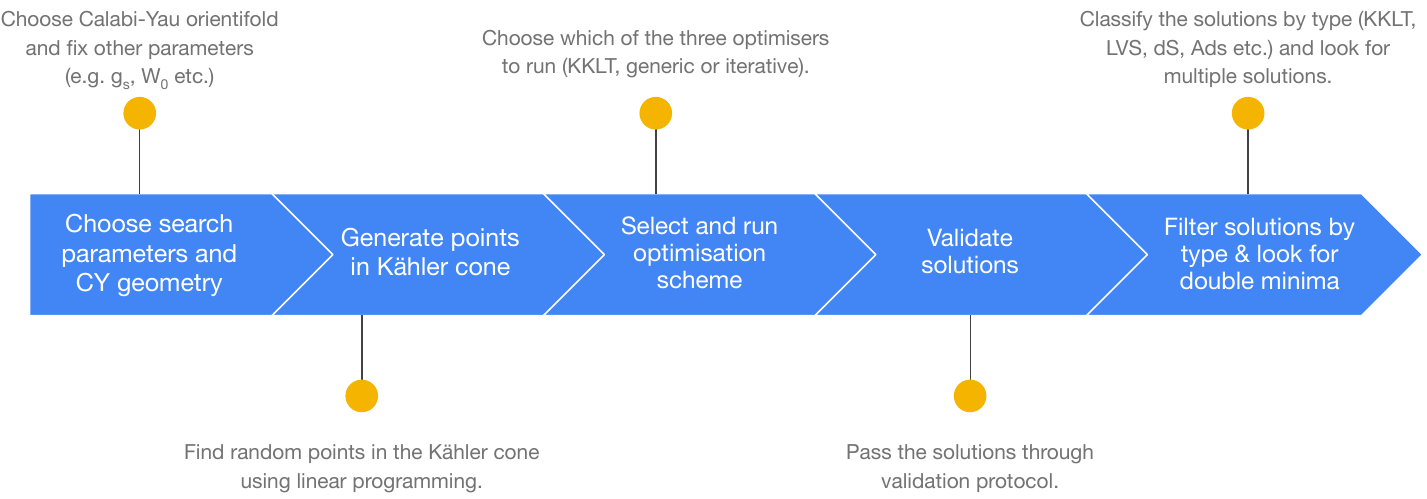}
    \caption{A schematic summary of the process followed to obtain string vacua numerically using the techniques and tests outlined in \S\ref{subsec:numeric-tests}.}\label{fig:numerical_pipline}
\end{figure}

In the section, we detail our computational framework for large scale numerical Kähler moduli stabilisation. The layout of our implementation is summarised in Fig.~\ref{fig:numerical_pipline}. First, we briefly summarise the status quo for numerical moduli stabilisation and lay out the plan and target for our implementation.

\subsection{Numerical Kähler moduli stabilisation}\label{sec:SOAnumerics}

The primary challenge in applying numerical methods to  string moduli stabilisation stems from the high dimensionality and complex functional form of the scalar potential given in equation \eqref{eq:master-eqn}. For a system with $h^{1,1}$ complex K\"ahler moduli $T_i$, we must search a $(2h^{1,1})$-dimensional field space parametrised by $(t_i,\theta_i)$. The task is basically a $(2h^{1,1})$-dimensional constrained optimisation problem involving huge systems of coupled, highly non-linear equations.

A first step towards systematically exploring moduli potentials with the help of numerical tools was taken in \cite{AbdusSalam:2020ywo} by working with an explicit implementation of \eqref{eq:master-eqn} (with $\alpha=0$). Hybrid optimisation algorithms and other computational techniques were employed to systematically explore scalar potentials in Calabi-Yau compactifications of Type IIB with $h^{1,1}\leq 3$ for minima. 

In principle, one could envision developing algorithms capable of handling hundreds of Kähler moduli, similar to those described previously \cite{Demirtas:2021nlu,McAllister:2024lnt}.\footnote{See also \cite{Gendler:2023hwg} for progress in finding axion minima.} While such implementations currently favour KKLT-like solutions, the minimisation algorithm of \cite{Demirtas:2021nlu} is capable of exploring the extended Kähler cone and, after suitable modifications, allows to include anti-D3-uplifts and backreaction from complex structure moduli \cite{McAllister:2024lnt}.
Moreover, by implementing detailed analytic expressions summarised in App.~D of \cite{McAllister:2024lnt}, all tree level $\alpha'$ corrections including worldsheet instanton contributions can be incorporated in the analysis.

\subsection{The implementation}\label{subsec:NT-reducing_computations_per_run}

In this work, we make progress by providing a versatile and efficient implementation to expedite the numerical search for string vacua. The $F$-term scalar potential is completely specified by the notion of the K\"ahler potential $K$, the holomorphic Kähler coordinates $T_i$ and the superpotential $W$. By providing only $K$, $W$ and $T_i$ as inputs, our code derives the full effective theory with the help of automatic differentiation. 

To achieve this, we are using the JAX  ecosystem \cite{jax2018github} which provides various features. By using auto-differentiation, it allows for efficient, and precise gradient computations in models with complex mathematical structures, such as those found in string compactifications. Compared to traditional means to compute derivatives like via finite differences or symbolic differentiation, this approach proves particularly powerful in enabling scalable explorations of high-dimensional parameter spaces. Building on this foundation, we have developed an algorithmic framework that generates all necessary functions to evaluate both the scalar potential and $F$-term conditions. These general-purpose functions remain model-agnostic until supplied with compactification data from the Calabi-Yau orientifold, including $\chi$ and $h^{1,1}$, and specific parameters from higher-scale physics (such as $g_s$, $K_{\text{cs}}$, and $W_0$).

A clear advantage is that our implementation is less prone to errors.
Previous implementations can only be used for a very specific ansatz for $K$, $W$ and $T_i$ making them susceptible to errors whenever significant modifications are required. Instead, our framework is completely modular allowing us to work with arbitrary inputs for $K$, $W$ and $T_i$. In this sense, our approach allows to explore the string landscape more broadly by modelling different forms of the perturbative corrections to $K$ (and $T_i$) as well as by being able to find arbitrary minima for Kähler moduli as explained in the subsequent section. In this work, we will work with the specific expressions \eqref{eq:BBHL+log-pert-corrections} and \eqref{eq:Wpot-defs} for $K$ and $W$.

Beyond these enhancements to physics capabilities, our implementation features increased computational speed. While for standard python implementations the computational cost of evaluating the scalar potential and its derivatives typically scales unfavourably with $h^{1,1}$, our algorithms alleviate this problem through \texttt{just-in-time} compilation as previously argued in \cite{Dubey:2023dvu} for flux vacua.

\subsection{Minimisation of scalar potentials and validation of solutions}
\label{subsec:numeric-tests}

To locate minima of the scalar potential, we employ three distinct numerical approaches. The first procedure directly solves the $F$-term conditions via Eq.~\eqref{eq:trivial-moduli-relation-ii}. The remaining two procedures are solution-agnostic, working with the full scalar potential, thereby enabling broader exploration of vacua; they differ primarily in how the parameter space is sampled.

We introduce $t_0^i$ as the initial guess for the two-cycle volumes. To sample $t_0^{i}$ inside the Kähler cone $\mathcal{K}_{X}$ of the Calabi-Yau threefold $X$, we impose the Kähler cone conditions in Eq.~\eqref{eq:Hineq} on the $t_0^i$ as a set of numerical inequalities. These are solved using linear programming via \texttt{scipy.optimize.linprog} to generate valid starting points $t_0^{i}\in \mathcal{K}_X$.

We employ the following three strategies to obtain a minimum of the scalar potential, whose associated Kähler parameters are denoted by $t_\star^i$:
\begin{enumerate}
    \item[1.] We solve the $F$-term conditions $D_iW=0$ numerically.
    As discussed in \S\ref{subsubsec:KKLT-results} and similar to \cite{Demirtas:2021nlu,McAllister:2024lnt}, the logarithmic form of these conditions in Eq.~\eqref{eq:trivial-moduli-relation-ii} offers significant computational advantages.
    \item[2.] We operate directly on the scalar potential trying to find solutions to the extremum equations $\partial_i V=0$. This allows us to locate non-supersymmetric minima of the Kähler moduli potential, including LVS-type solutions.\footnote{Of course, a subset of solutions obtained in this way may still preserve supersymmetry due to $D_iW=0$ as we will see later in \S\ref{subsec:known-solns-results}.}
    \item[3.] By again working with the full scalar potential, we employ the iterative method described in \S\ref{subsubsec:Westphal-derv} by first locating KKLT solutions and then using these as initial conditions to search for non-SUSY vacua at incrementally larger values of $|W_0|$. This approach proves effective in identifying K\"ahler uplifted solutions that would otherwise be inaccessible via direct minimisation.\footnote{Similar strategies were employed in \cite{McAllister:2024lnt} to find de Sitter minima from anti-D3-brane uplifts.}
\end{enumerate}
For each strategy, we solve the resulting system of equations via \texttt{scipy.optimize.root}. For each candidate solution, we check that the minimum lies inside Kähler cone $t_\star^i\in \mathcal{K}_X$ and that the Hessian eigenvalues are positive.\footnote{In principle, we could allow for some negative eigenvalues in AdS minima provided that the tachyonic mass is above the Breitenlohner-Freedman (BF) bound \cite{Breitenlohner:1982jf}.}

Beyond these minimisation methods, we strategically choose initial points $t_0^{i}\in \mathcal{K}_X$ to optimise convergence.  E.g., in two-modulus LVS scenarios, prior knowledge guides the selection of starting points: the large modulus typically stabilises near $\tau_b \sim \mathcal{V}^{2/3}$ and the small modulus satisfies $a_s \tau_s \sim \ln(\mathcal{V})$, cf. Eq.~\eqref{eq:LVS_solution}. By finding the corresponding Kähler parameters $t_0^{i}\in \mathcal{K}_X$ where these conditions are approximately satisfied, we increase the likelihood of finding LVS solutions, even in examples with $h^{1,1}>2$.

\section{Numerical landscape analysis and coexisting minima}
\label{sec:results}

In this section, we present the numerical results obtained using the computational framework described in the previous section.
First, we describe our selection of datasets in \S\ref{sec:initscan}. Subsequently, we demonstrate in \S\ref{subsec:known-solns-results} that applying in our numerical techniques successfully recovered all known minimum types described in \S\ref{subsec:known-solns-derv} across all sampled geometries with $2\leq h^{1,1}\leq 6$. We also perform more in-depth searches in \S\ref{subsec:known-solns-results} on individual manifolds to highlight the prevalence and parametric distributions of certain string vacua within our datasets.

A particularly interesting outcome of our numerical searches is the generic appearance of double minima candidates in manifolds supporting multiple solution types. As detailed in \S\ref{subsec:double-mins}, we consistently find pairs of minima --- including AdS-AdS, Minkowski-AdS, and dS-AdS configurations --- existing simultaneously within the same potential (i.e., for identical values of parameters such as $\chi$, $g_s$, $W_0$, etc.). We comment on the broader implications of vacuum stability and inter-minima tunnelling in \S\ref{sec:P11169}. In \S\ref{subsec:h11-3+_and_uplift}, we extend our analysis beyond the $F$-term scalar potential by including explicit uplift terms, particularly from anti-D3 branes. These uplift mechanisms allow us to convert AdS vacua into metastable de Sitter vacua.

\subsection{Search parameters and dataset selection}\label{sec:initscan}

\begin{table}[t!]
    \centering
    \begin{tabular}{c|c c c}
       $\numod$  & $\#$ Geometries & $\%$ Success & Time per run (s)\\
        \hline
      2 & 39    
      & 21\%     & 0.95s \\
      3 & 274
      & 6.4\%    & 1.2s \\
      4 & 1,760
      & 2.2\%    & 2.0s\\
      5 & 11,713
      & 0.3\%    & 2.9s\\
      6 & 74,503 
      & 0.3\%    & 4.0s \\
    \end{tabular}
    \caption{Table summarising the performance of the algorithm over of our shallow scan over all geometries in the database of \cite{Crino:2022zjk} with $h^{1,1}\leq6$ where optimisation was attempted for 18 combinations of $(g_s,W_0)$ per geometry. It should be noted that `time per run' refers to the average time\protect\footnotemark~taken to conduct the first 4 of 5 steps in Fig.~\ref{fig:numerical_pipline}, from generating a start point in the Kähler cone to running all three optimisation methods and passing any solutions found through our checks.}\label{tab:large_scan}
\end{table}
\footnotetext{The average time was taken for the first 1,000 geometries for $\numod\geq 4$ rather than the whole database for efficiency.}

Initially, we performed large scale searches across the toric Calabi-Yau database of \cite{Crino:2022zjk} for $h^{1,1}\leq 6$. Specifically, we run superficial scans using the three approaches described in \S\ref{subsec:numeric-tests} using three values of $g_s \in \{0.03,0.1,0.3\}$ and six values of $-W_0 \in [10^{-15}, 10^{0.5}]$, for a total of 18 parameter points per geometry. Furthermore, we assume all $h^{1,1}$ K\"ahler moduli to contribute to the non-perturbative superpotential.\footnote{Although this has been explicitly confirmed for large values of $h^{1,1}$, for smaller values it must be verified on a case-by-case basis. In this sense, our analysis applies only to cases where the $h^{1,1}$ $4$-cycles are rigid or can be rigidified.} Geometries were then ranked by their success rate in this quick scan, guiding our selection for more detailed, fine-grained searches described in \S\ref{subsec:known-solns-results}. The computational efficiency of our approach is demonstrated by successful identifications of both SUSY and non-SUSY minima, with typical runtimes of just a few seconds on standard laptop hardware. We summarise the number of geometries per $h^{1,1}$, the average success rate for finding vacua as well as the average time per run in Table~\ref{tab:large_scan}.\footnote{We use the counting of FRST classes of \cite{Gendler:2023ujl}. We note that, while these Calabi-Yau threefolds have distinct Wall data, they can still be topologically equivalent as complex manifolds. Topological inequivalence classes of Calabi-Yau hypersurfaces constructed from triangulations of reflexive polytopes in four dimensions for $h^{1,1}\leq 5$ were enumerated in \cite{Gendler:2023ujl}, see also \cite{Chandra:2023afu} for related attempts.}

\begin{figure}[t!]
    \centering
    \includegraphics[scale=0.48]{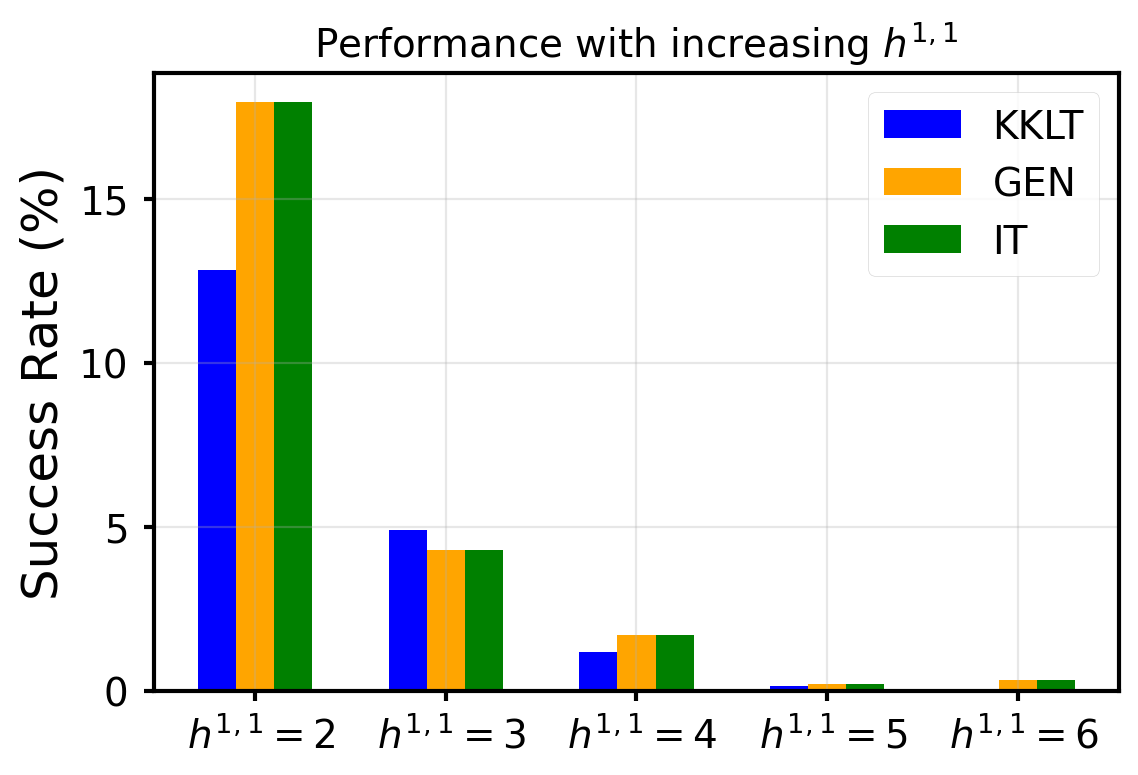}
    \hspace*{0.3cm}
    \includegraphics[scale=0.48]{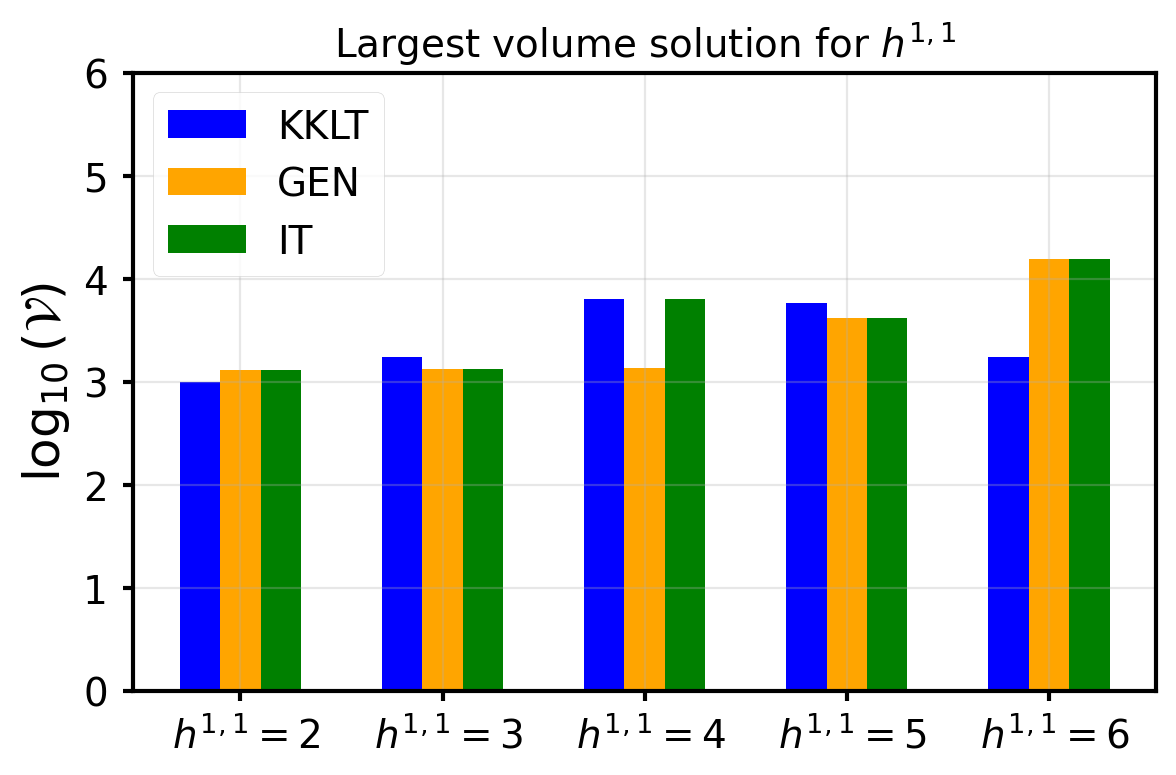}
    \caption{Success rate (left) and largest volume solution (right) for the large scan over the geometries listed in Table~\ref{tab:large_scan}. Here, the different colours correspond to the individual minimisation strategies as explained in \S\ref{subsec:numeric-tests}: blue for supersymmetric, yellow for general non-supersymmetric and green for solutions found via the iterative procedure. We observe that the success rate of all strategies decreases with $h^{1,1}$, whereas the largest volume obtained in any of the solutions stays roughly constant.
    }
    \label{fig:Altman_scan_h11_plots}
\end{figure}

Fig.~\ref{fig:Altman_scan_h11_plots} shows the success rate for the three different minimisation strategies separately as well as the largest volume solution obtained for each $h^{1,1}$. The drop in success rate for uncovering minima per starting point is a natural consequence of our coarse initial scan and can be substantially improved with more targeted, in-depth searches. Indeed, this preliminary survey was designed to pinpoint the most promising geometries for the more detailed investigations that follow below. Specifically, we next demonstrate that our framework, detailed in \S\ref{sec:numerics}, can be efficiently used to recover all established solution types described in \S\ref{subsec:known-solns-derv}, namely KKLT, LVS, and K\"ahler uplifted solutions, as well as hybrid solutions as found previously in \cite{AbdusSalam:2020ywo}.

\subsection{Numerical KKLT, LVS and Kähler uplift solutions}
\label{subsec:known-solns-results}

We now present solutions that are anticipated based on the analytic considerations of \S\ref{subsec:known-solns-derv}, before summarising novel solution types subsequently.

\subsubsection*{KKLT solutions}
\label{subsubsec:KKLT-results}

The KKLT minimisation method proves particularly efficient, consistently outperforming direct minimisation of the full scalar potential by solving the simpler $F$-term conditions directly. A key advantage is the usage of the logarithmic expression \eqref{eq:trivial-moduli-relation-ii} for the Kähler moduli $T_i$.

\begin{figure}[t!]
    \centering
    \includegraphics[width = \linewidth]{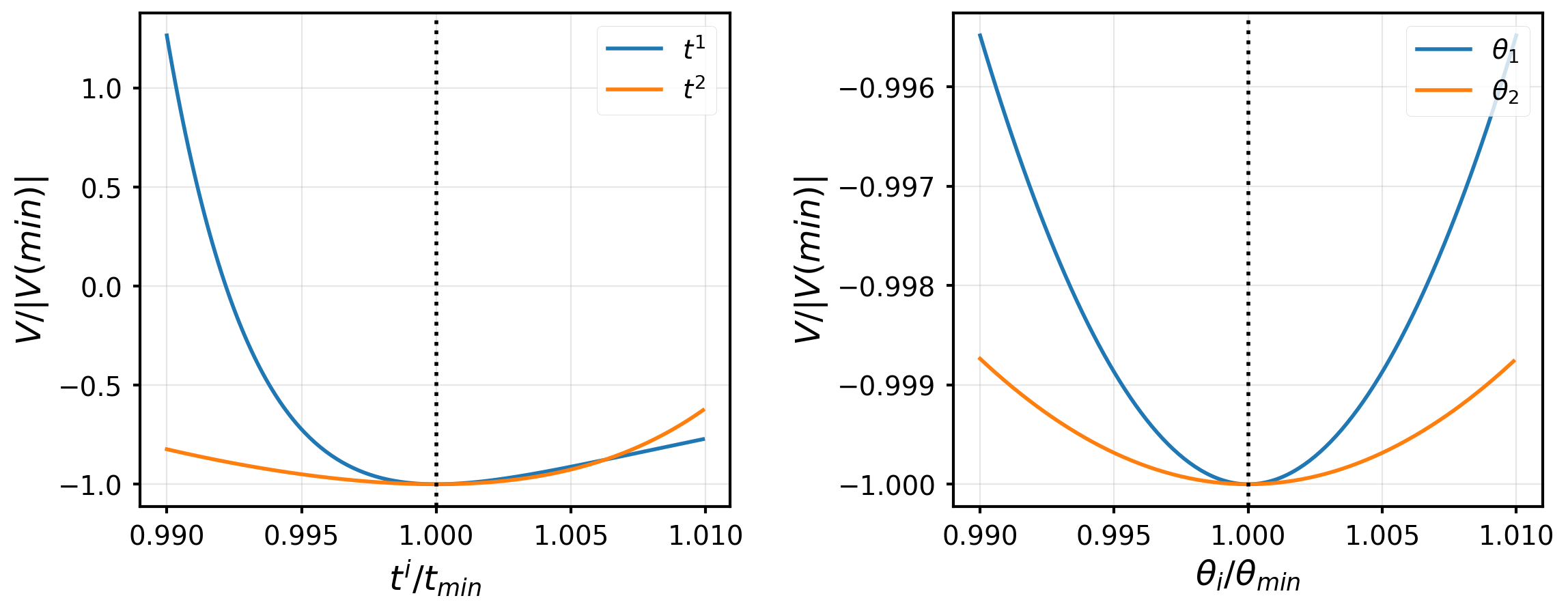}
    \caption{Example SUSY AdS minimum obtained by solving the $F$-term conditions for a potential generated with parameters as specified in Eq.~\eqref{eq:KKLTEx1Par}.}\label{fig:KKLT-example}
\end{figure}

Fig.~\ref{fig:KKLT-example} illustrates a representative KKLT minimum identified for compactifications on the degree 18 hypersurface in $\mathbb{P}^4_{[1,1,1,6,9]}$ by our $F$-term solver and subsequently verified through direct scalar potential minimisation. 
The potential is generated for parameters
\begin{equation}\label{eq:KKLTEx1Par}
    g_s=10^{-1.5}\, ,\; W_0 = -10^{-12}\, ,\; A_1=A_2=1\, ,\; a_1=\frac{\pi}{24}\, ,\;a_2=\frac{\pi}{22}\, , \; \alpha=10^{-1} .
\end{equation}
The calculation incorporates both BBHL and logarithmic K\"ahler potential corrections, with $\alpha=10^{-1}$, yielding a SUSY minimum at 
\begin{equation}
    \mathcal{V}^{(0)} \approx 1{,}410\, ,\;\qquad \dfrac{\delta \mathcal{V}_{\text{BBHL}}}{\mathcal{V}^{(0)}}\sim 10^{-1}\, ,\;\qquad \dfrac{\delta \mathcal{V}_{\text{log}}}{\mathcal{V}^{(0)}}\sim -10^{-2}\, .
\end{equation}
The vacuum energy is given by
\begin{equation}
    \langle V\rangle \approx -5.7\times10^{-34}\, .
\end{equation}

\begin{figure}[t!]
    \centering
     \includegraphics[scale=0.45]{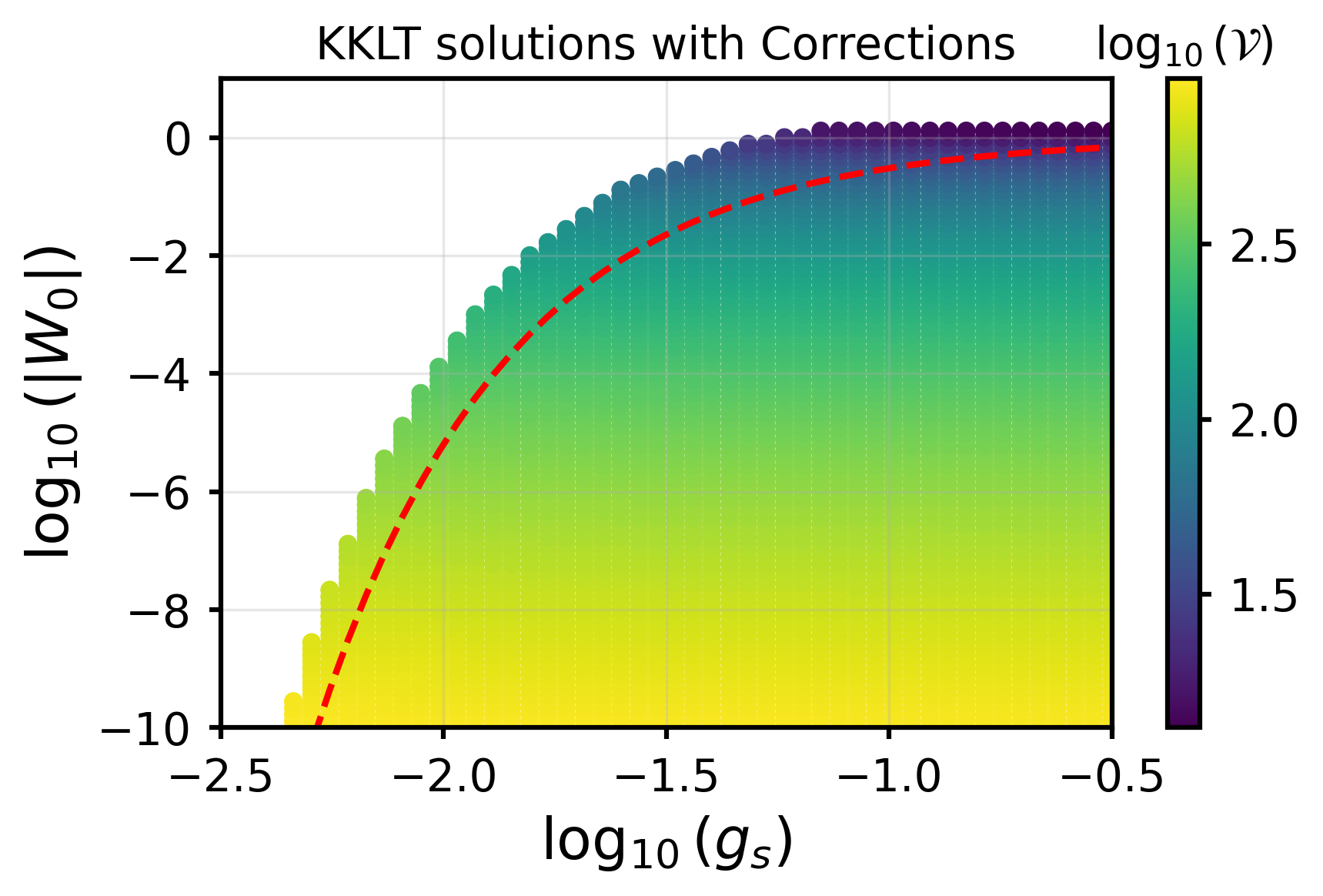}
     \hspace{0.5cm}
     \includegraphics[scale=0.45]{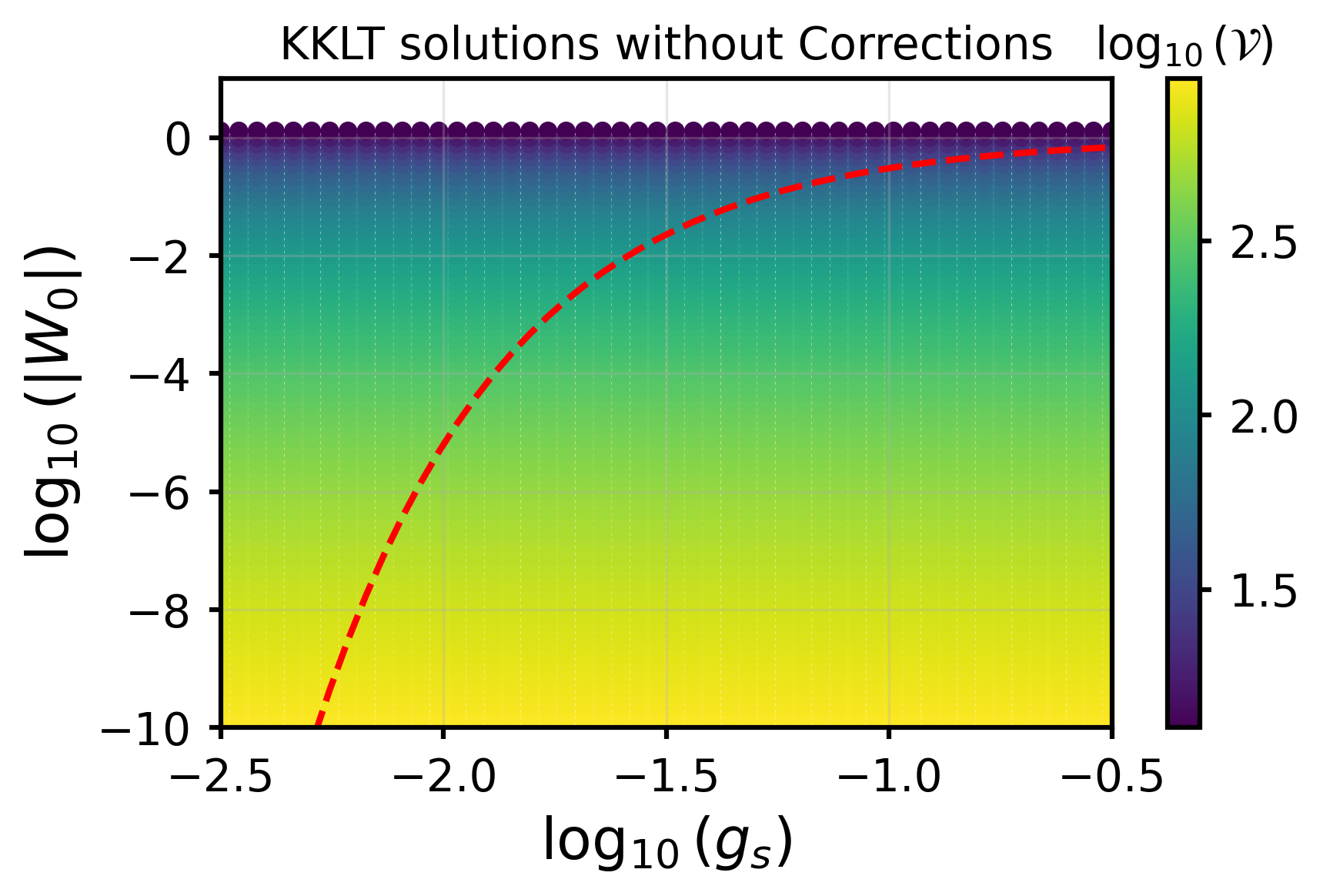}
    \caption{Solution space obtained from solving the $F$-term conditions (Eq.~\eqref{eq:trivial-moduli-relation-ii}) across the parameter ranges of $g_s$ and $|W_0|$ described in the text. Left panel displays results without perturbative K\"ahler potential corrections, while right panel includes both BBHL and logarithmic corrections (Eq.~\eqref{eq:BBHL+log-pert-corrections}). Data obtained for Calabi-Yau manifold from polytope with vertices \eqref{eq:poly52} ($h^{1,1}=3$) using parameters: $A_i=1$, $a_i=2\pi/22$, Euler characteristic $\chi=-112$, and logarithmic correction coefficient $\alpha=10^{-1}$. 
    }\label{fig:KKLT-dataset}
\end{figure}

As a representative case study, Fig.~\ref{fig:KKLT-dataset} presents results from scanning a CY manifold with $\numod=3$ whose polytope is specified by the following vertices
\begin{equation}\label{eq:poly52}
    \left(\begin{array}{ccccccc}
    \phantom{-}1& -3& -2& -2& \phantom{-}0& \phantom{-}0& \phantom{-}1\\
    \phantom{-}0&  -1& -1& \phantom{-}0& \phantom{-}0 & \phantom{-}1 & \phantom{-}1\\ 
    \phantom{-}0&  -1&  \phantom{-}0&  -1& \phantom{-}1 & \phantom{-}0 & \phantom{-}1 \\  
    \phantom{-}0&  -2&  \phantom{-}0&  \phantom{-}0& \phantom{-}0 & \phantom{-}0 & \phantom{-}2\\
    \end{array}\right)\,.
\end{equation}
The triangulation is determined by the height vector $(-1,  0,  3,  3,  0,  0,  0,  0)$. As can be verified explicitly, this geometry has more than $h_{11}$ rigid divisors (see e.g. \cite{Crino:2022zjk}).
Our search employed the $F$-term conditions directly, exploring 50 values of $g_s$ in $[10^{-2.5}, 10^{0.5}]$ and 100 values of $|W_0|$ in $[10^{-10}, 10^{1}]$ totalling 5,000 individual runs.

Fig.~\ref{fig:KKLT-dataset} displays the parameter space regions where KKLT solutions exist, with coloured points indicating solutions and colours representing the stabilised volume. As anticipated for KKLT solutions, the volume depends strongly on $|W_0|\sim A\mathrm{e}^{-a\tau}$ but remains largely insensitive to the value of $g_s$. This behaviour follows from Eq.~\eqref{eq:trivial-moduli-relation-ii}, where $g_s$ influences K\"ahler moduli stabilisation only through subleading perturbative corrections to the K\"ahler potential. Consequently, the inclusion of $g_s$-dependent perturbative corrections produces negligible changes in the stabilised volumes.

Comparing scans with and without perturbative corrections shows that the presence of $g_s$-dependent corrections shrink the viable $|W_0|$ window for KKLT minima. 
In particular, the BBHL correction scales as $g_s^{-3/2}$, pushing viable vacua to ever smaller $|W_0|$ as $g_s$ decreases. As discussed in \S\ref{subsubsec:Westphal-derv}, the full scalar potential (cf.~Eq.~\eqref{eq:master-eqn}) develops singularities when $\mathcal{V} \sim \delta \mathcal{V}(t^i,s)$. For KKLT solutions, the volume scaling $a_i\, \mathrm{Re}(T_i) \sim \ln(|W_0|^{-1})$ (recall Eq.~\eqref{eq:trivial-moduli-relation-ii}) motivates the dotted line in each plot showing $\ln(|W_0|^{-1}) = \chi^{2/3}a_1/g_s$, which approximates where $\mathcal{V} \approx \delta \mathcal{V}_{\text{BBHL}}$ and correctly predicts the observed scaling behaviour.

\subsubsection*{LVS solutions}
\label{subsubsec:LVS-results}

Next, let us employ our search algorithm to identify LVS minima by working directly with the full scalar potential in Eq.~\eqref{eq:master-eqn}. That is, instead of solving $D_{T_i}W=0$ as in the previous section, we are searching for solutions to $\partial_{t^i}V=\partial_{\theta_i}V=0$ from random starting points $t_0^i$ within the K\"ahler cone $\mathcal{K}_X$. A priori, this may not automatically give rise to an LVS-type solution, but simply to a generic (non-)SUSY (A)dS$_4$ minimum. Locating such non-SUSY AdS$_4$ minima therefore presents significantly greater challenges than their supersymmetric counterparts. 

\begin{figure}[t!]
    \centering
    \includegraphics[width=\linewidth]{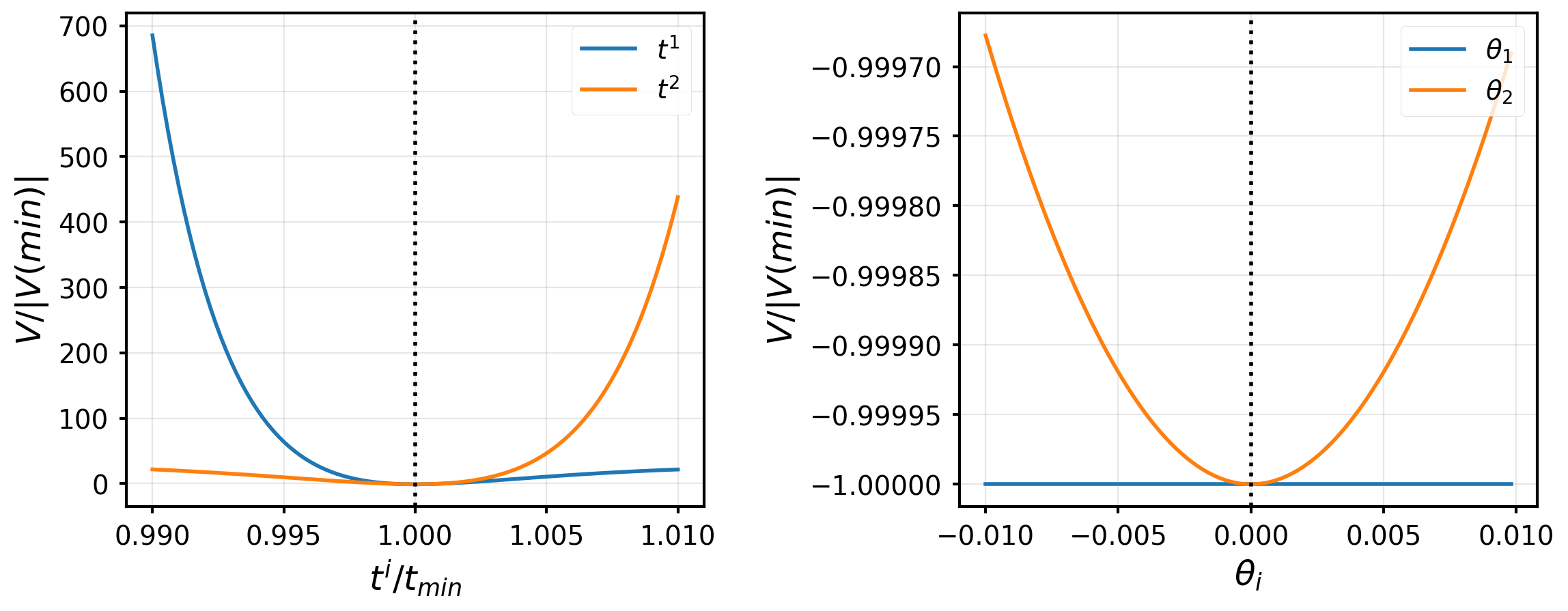}
    \caption{LVS minimum identified through direct minimisation of the scalar potential with parameters as specified in Eq.~\eqref{eq:LVSEx1Par}.}
    \label{fig:LVS-example}
\end{figure}

We present one representative minimum at $h^{1,1}=2$ in Fig.~\ref{fig:LVS-example}. This minimum shows the expected characteristics of LVS solutions.
As before, we consider compactifications on $\mathbb{P}^4_{[1,1,1,6,9]}[18]$ with both BBHL and logarithmic perturbative corrections.
The minimum was obtained for the parameters
\begin{equation}\label{eq:LVSEx1Par}
     g_s=0.1\, , \; W_0 = -0.628\, , \; A_1=A_2=1\, , \; a_1=a_2=2\pi/20\, , \; \alpha=10^{-1} \, .
\end{equation}
The minimum occurs at 
\begin{equation}
    \mathcal{V}^{(0)} \approx 23{,}400\, ,\;\qquad \dfrac{\delta \mathcal{V}_{\text{BBHL}}}{\mathcal{V}^{(0)}} \sim 10^{-3}\, ,\;\qquad \dfrac{\delta \mathcal{V}_{\text{log}}}{\mathcal{V}^{(0)}} \sim -10^{-3}
\end{equation}
with vacuum energy 
\begin{equation}
    \langle V\rangle \approx -4.9\times10^{-15}\, .
\end{equation}
The solution exhibits characteristic LVS scaling as derived in Eq.~\eqref{eq:LVS_solution}. That is, the blow-up and volume modulus satisfy
\begin{equation}
    \tau_s = \tau_2 \approx 31.4 \approx 0.98\, a_s^{-1}\ln(\mathcal{V})\, ; \,\qquad \tau_b = \tau_1 \approx 927 \approx 1.21 \, \mathcal{V}^{2/3}\, .
\end{equation}

\begin{figure}[t!]
    \centering
     \includegraphics[width=\linewidth]{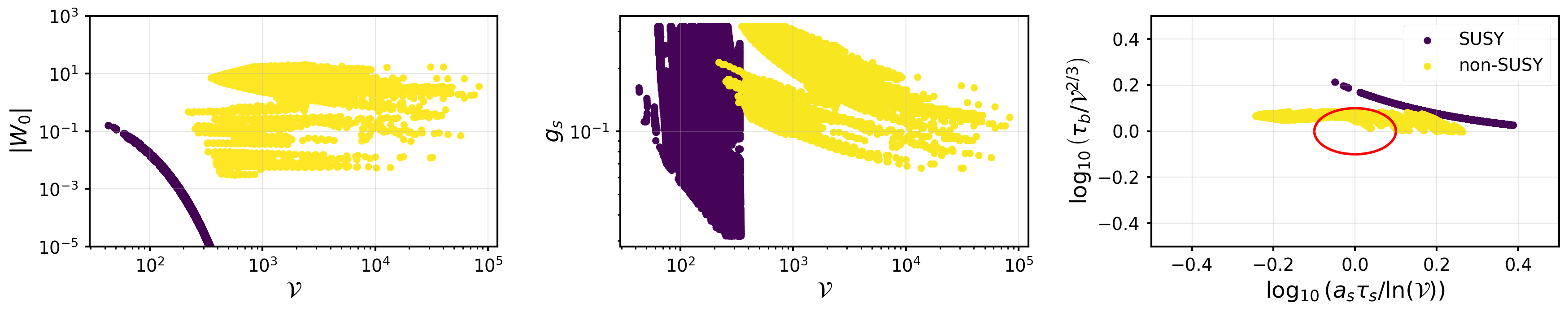}
    \caption{
    Solution space obtained by directly minimising the full potential for compactifications on $\mathbb{P}^4_{[1,1,1,6,9]}[18]$ ($h^{1,1}=2$) with $A_i=1$, $a_i=2\pi/20$, $\chi=-540$, and $\alpha=10^{-1}$.  In total, we show 19{,}484 vacua found (including KKLT-type solutions) containing 2{,}750 LVS minima.
    }
    \label{fig:LVS-dataset}
\end{figure}

Our numerical algorithms successfully generate comprehensive LVS solution datasets, revealing characteristic parameter scalings. Fig.~\ref{fig:LVS-dataset} presents results from scanning 50,000 parameter combinations across $g_s \in [0.03,0.3]$ and $|W_0| \in [10^{-5}, 10^{2}]$. By minimising the full potential, we see that our method finds a large majority of supersymmetric solutions. This is not surprising as the LVS solutions lie in asymptotic regions of the moduli space and are therefore much harder to find.\footnote{In principle, one can devise methods to find approximate LVS solutions: starting from some point $t^i_0\in\mathcal{K}_X$, locate points $t^i_1\in\mathcal{K}_X$ in Kähler moduli space $\tau_k\sim \ln(\mathcal{V})$ for some $k=1,\ldots,h^{1,1}$ and subsequently use $t^i_1\in\mathcal{K}_X$ as the starting point for finding the true minimum of the potential. Similar ideas were employed in \cite{Demirtas:2021nlu,McAllister:2024lnt} to find KKLT minima.}
After filtering out supersymmetric solutions, we obtain the plot on the right hand side in Fig.~\ref{fig:LVS-dataset}.
This plot shows that, as we vary $|W_0|$ and $g_s$, the lower volume LVS solutions coalesces with KKLT-like solutions, which are growing in volume. This is a hint that, for a single choice of parameters like $W_0$ and $g_s$, the same potential can host both LVS- and KKLT-like minima. This will be explored further in \S\ref{subsec:vary-W0-results}.

Much as in Fig.~\ref{fig:KKLT-dataset}, the first two panels of Fig.~\ref{fig:LVS-dataset} map where vacua appear in the $(|W_0|,g_s)$-plane.  Unlike the KKLT case where $\mathcal{V}$ depends almost exclusively on $\lvert W_0\rvert$, LVS minima show a two‑parameter sensitivity: volumes decrease as $|W_0|$ is lowered but grow for smaller $g_s$.  This behaviour matches the characteristic LVS scaling $\mathcal{V}\sim g_s^{-1/2}|W_0|\,\mathrm{e}^{1/g_s}$, and explains why LVS vacua persist to much larger $|W_0|$ than their KKLT counterparts, even overlapping in parameter space, see \S\ref{subsec:double-mins} and the stability discussion in \S\ref{subsec:tunnelling}. 

The third plot in Fig.~\ref{fig:LVS-dataset} demonstrates our LVS check from \S\ref{subsec:numeric-tests}. Here, we test how close our solutions come to satisfying the standard LVS relations $a_s \tau_s \sim \ln(\mathcal{V})$ and $\tau_b \sim \vol^{2/3}$. We thereby identify valid LVS solutions within a defined region around the origin.

\subsubsection*{K\"ahler uplifted minima}
\label{subsubsec:small-vol-non-susy-results}

Next, let us describe our results for the third class of solutions, namely Kähler uplifted minima \cite{Balasubramanian:2004uy,Westphal:2006tn} as reviewed in \S\ref{subsubsec:Westphal-derv}.
To reiterate, these solutions emerge when $|W_0|$ exceeds a certain threshold where SUSY-breaking effects from $\alpha'$-corrections provide a source for uplifting.

\begin{figure}[t!]
    \centering
    \includegraphics[width=\linewidth]{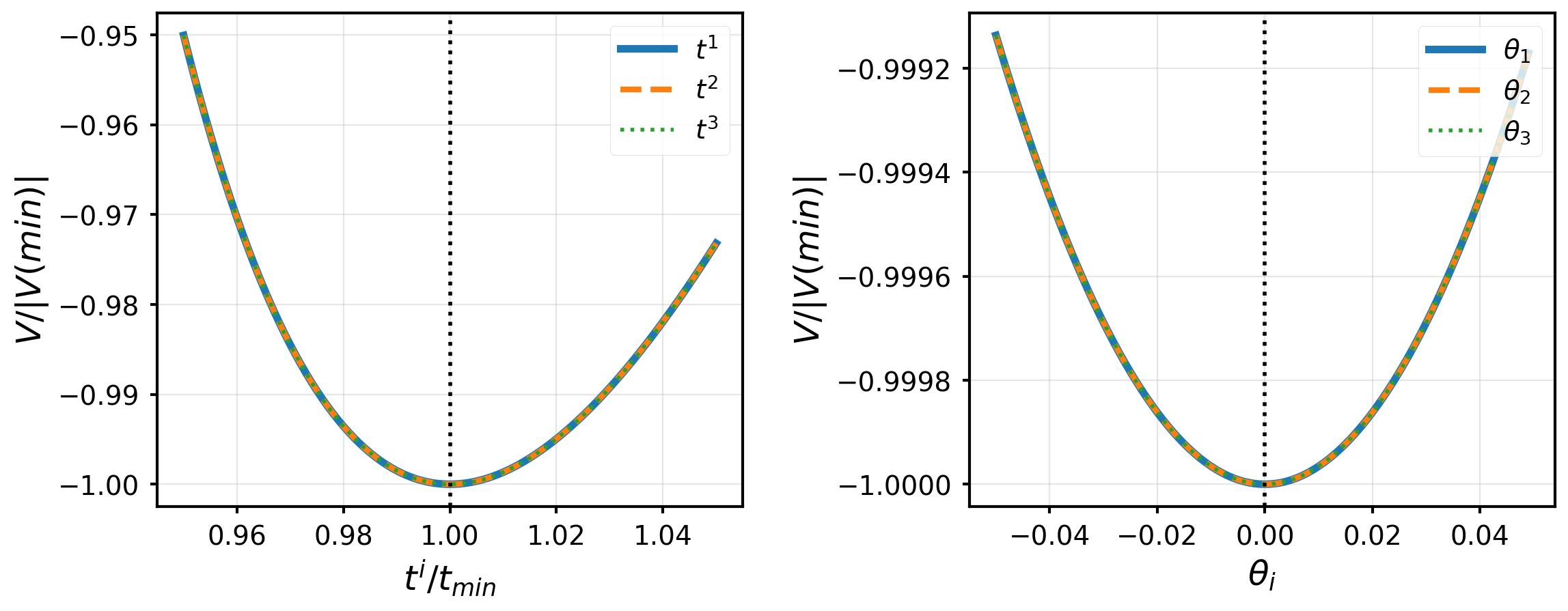}
    \caption{K\"ahler uplifted AdS minimum obtained for the Calabi-Yau hypersurface at $h^{1,1}=3$ obtained from the polytope defined by \eqref{eq:poly52}. The parameters for this minimum are given in Eq.~\eqref{eq:KupliftEx1Par} }\label{fig:west_example}
\end{figure}

Our search algorithm employs an iterative approach: beginning with a known KKLT minimum at small $|W_0|$, we gradually increase $|W_0|$ while using each solution as the initial guess for subsequent optimisations for $\partial_{t^i}V=\partial_{\theta_i}V=0$.
This incremental parameter variation combined with informed starting points enables efficient tracking of solutions to non-SUSY AdS$_4$ minima and potential uplift to dS$_4$ vacua at shrinking volumes. Fig.~\ref{fig:west_example} illustrates a representative example of a non-SUSY AdS$_4$ minimum found for the toric Calabi-Yau hypersurface with polytope determined by the vertices in \eqref{eq:poly52} which has $\chi=-112$.
We obtained a minimum for the following choice of parameters
\begin{equation}\label{eq:KupliftEx1Par}
    g_s=0.0703\, ,\; W_0=-1.23\, ,\; A_1=A_2=A_3=1\, ,\; a_1=a_2=a_3=\dfrac{2\pi}{22}\, ,\; \alpha=10^{-1}\, .
\end{equation}
The minimum occurs at volume
\begin{equation}\label{eq:KupliftEx1Vol}
    \mathcal{V}^{(0)}\approx 19.7\, ,\;\qquad \dfrac{\delta \mathcal{V}_{\text{BBHL}}}{\mathcal{V}^{(0)}} \approx 0.5\, ,\;\qquad \dfrac{\delta \mathcal{V}_{\text{log}}}{\mathcal{V}^{(0)}} \sim  -10^{-2}\, ,
\end{equation}
yielding a value for the vacuum energy
\begin{equation}
    \langle V\rangle \approx -1.23\times10^{-4}\, .
\end{equation}
Clearly, looking at the size of the volume and corrections in \eqref{eq:KupliftEx1Vol}, this vacuum most likely lies outside the regime of perturbative control.

\begin{figure}[t!]
    \centering
    \includegraphics[width=\linewidth]{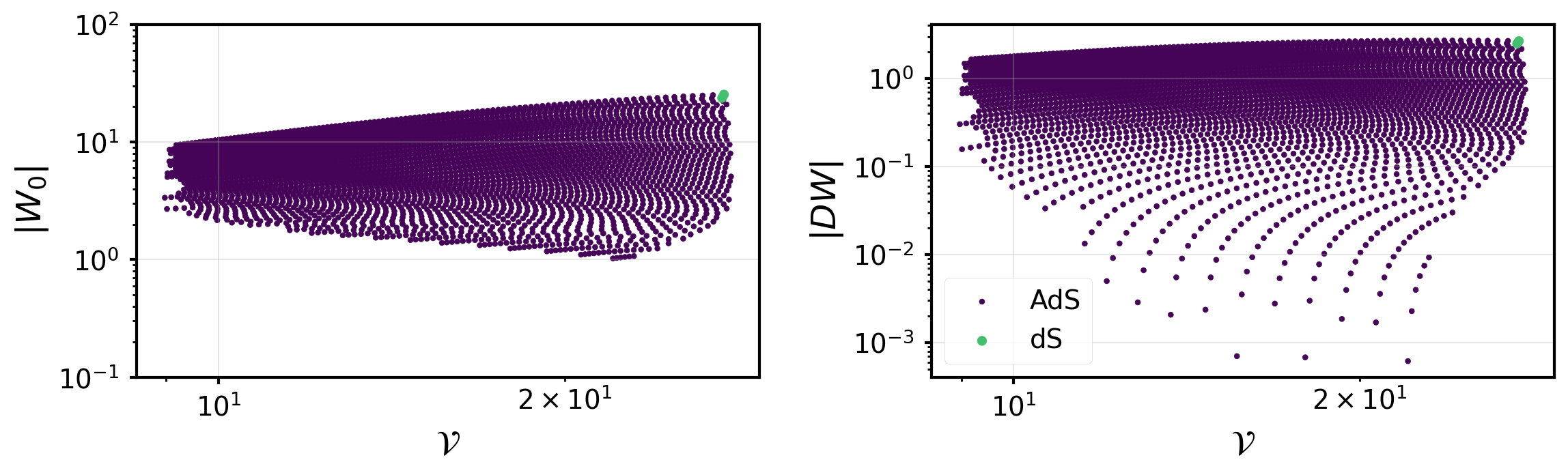}
    \caption{
    We obtained 2{,}864 K\"ahler uplifted minima across the scanned parameter ranges of $g_s$ and $|W_0|$ with de Sitter minima singled out in the plot on the right. 
    The data is obtained for the same Calabi-Yau threefold $h^{1,1}=3$ as in Fig.~\ref{fig:KKLT-dataset} and parameters $A_i=1$, $a_i=2\pi/22$, $\chi=-112$, and $\alpha=10^{-1}$.
    }
    \label{fig:west-dataset}
\end{figure}

K\"ahler uplifted solutions emerge at larger $|W_0|$ values where $F$-term minimisation cannot be achieved within the regime of perturbative control. To map this transition, we conduct parallel scans for both KKLT and non-SUSY solutions across identical parameter ranges. 

Fig.~\ref{fig:KKLT-dataset} previously established the existence of a critical $|W_0|$ threshold (marked by the red dotted line at $\mathcal{V}^{(0)} \sim \delta\mathcal{V}$) beyond which KKLT solutions vanish. Using our iterative minimisation approach on the complete potential, we successfully extend the solution space beyond this boundary. The resulting K\"ahler uplifted minima, stabilised at smaller volumes through SUSY-breaking uplift effects, are presented in Fig.~\ref{fig:west-dataset}.

These K\"ahler uplifted minima exhibit several distinctive features.
First, these solutions occupy parameter regions with larger $|W_0|$ values than their KKLT counterparts, corresponding to smaller volumes. However, unlike KKLT solutions where volume decreases monotonically with $|W_0|$, the K\"ahler uplifted solutions show volume growth as $|W_0|$ increases further. Second, the volume of the solutions is small with $\mathcal{V}^{(0)}\sim 10$ signalling a probable loss of perturbative control.

Perturbative corrections establish an upper $|W_0|$ bound for KKLT solutions, beyond which K\"ahler uplifted solutions emerge in narrow $W_0$ bands for each $g_s$. These bands represent the delicate balance where corrections provide sufficient uplift without inducing runaway decompactification. The required parameter tuning reflects the KKLT solution's weak sensitivity to perturbative effects - in some cases, like the CY manifold of Fig.~\ref{fig:west-dataset}, the potential transitions directly to runaway before achieving positive values.

\begin{figure}[t!]
    \centering
    \includegraphics[width=\linewidth]{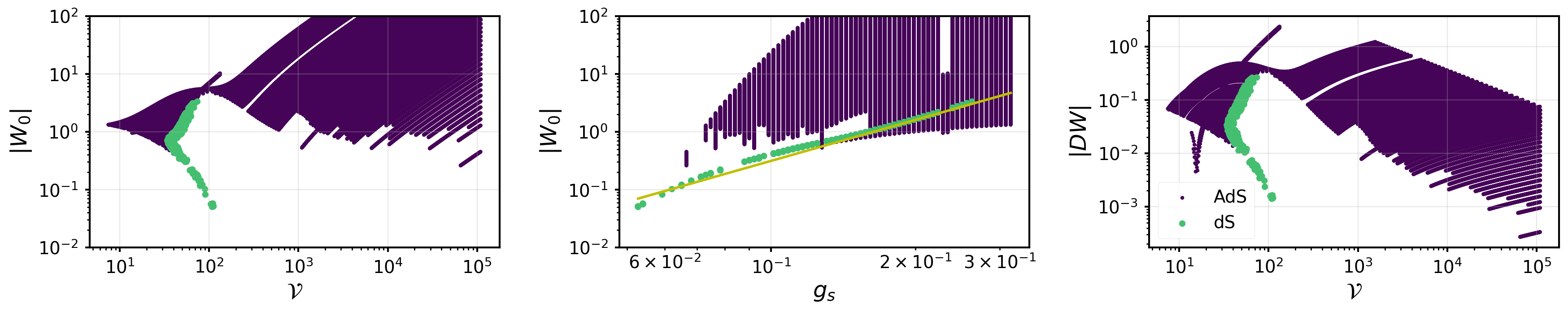}
    \caption{De Sitter solutions from K\"ahler uplifting in the small-volume non-SUSY regime. A subset of solutions to Eq.~\eqref{eq:master-eqn} yield positive potential values, confirming the prediction of \cite{Rummel:2011cd} for the Calabi-Yau threefold $\mathbb{P}^4_{[1,1,1,6,9]}[18]$ at $h^{1,1}=2$. The yellow curve shows their proposed relation $|W_0| \sim 70.2\, g_s^{2.35}$ where dS minima are expected to occur. For this plot, we use the same parameters as \cite{Rummel:2011cd}: $A_i=\{1.11,1\}$, $a_i=\{2\pi/24, 2\pi/22\}$, $\chi=-540$, $\mathrm{e}^{K_{\text{cs}}}=0.03$, and $\alpha=0$.
    }
    \label{fig:ds_west-dataset}
\end{figure}

The most interesting feature of these solutions is their ability to stabilise all moduli in de Sitter configurations through pure K\"ahler uplifting, without explicit uplift terms. Fig.~\ref{fig:ds_west-dataset} validates the prediction of \cite{Rummel:2011cd} that solutions along specific $g_s$-$W_0$ trajectories (yellow line) receive sufficient K\"ahler uplift to produce small-volume de Sitter minima.
More specifically, \cite{Rummel:2011cd} provide a condition on the value of the flux-induced superpotential and the topological data of the Calabi–Yau threefold that, if satisfied, ensures the existence of a metastable de Sitter vacuum. Indeed, working with the same choice of parameters as in \cite{Rummel:2011cd} for the Calabi-Yau threefold $\mathbb{P}^4_{[1,1,1,6,9]}[18]$ at $h^{1,1}=2$, our numerical analysis verifies that de Sitter vacua indeed arise along the expected trajectory $|W_0| \sim 70.2\, g_s^{2.35}$, see Fig.~2~in~\cite{Rummel:2011cd}.

\subsection{Additional hybrid minima}
\label{subsec:hybrid_solutions}

Throughout \S\ref{sec:EFT-and-soln-derivations} and \S\ref{sec:results}, we have organised vacua into three well-known classes, namely KKLT, LVS, and Kähler uplifted solutions, even though our numerical minimiser from \S\ref{sec:numerics} is agnostic about which type of solution it finds. One might therefore expect \emph{hybrid} minima outside these categories, as reported in \cite{AbdusSalam:2020ywo}.\footnote{We stress that our categorisation is slightly different to the one used in \cite{AbdusSalam:2020ywo}.} Such solutions can arise whenever moduli are stabilised e.g. through a combination perturbative corrections to the Kähler potential and non-perturbative corrections to the superpotential, but in the absence of a swiss-cheese like structure of the volume. More specifically, extending the LVS mechanism to $h^{1,1}>2$ generally demands additional perturbative or non-perturbative ingredients.  Whenever one modulus, say $\tau_b$, satisfies $\tau_b\sim\mathcal{V}^{2/3}$ and is stabilised by perturbative corrections, with the remaining small cycles fixed non-perturbatively, one recovers a genuine LVS vacuum.  In geometries without a clear swiss-cheese structure, or with multiple cycles receiving leading-order perturbative stabilisation, one instead finds hybrid minima that behave qualitatively like LVS vacua. Solutions of this type were coined \emph{LVS-like} hybrid minima in \cite{AbdusSalam:2020ywo}.

\begin{figure}[t!]
    \centering
    \includegraphics[width=\linewidth]{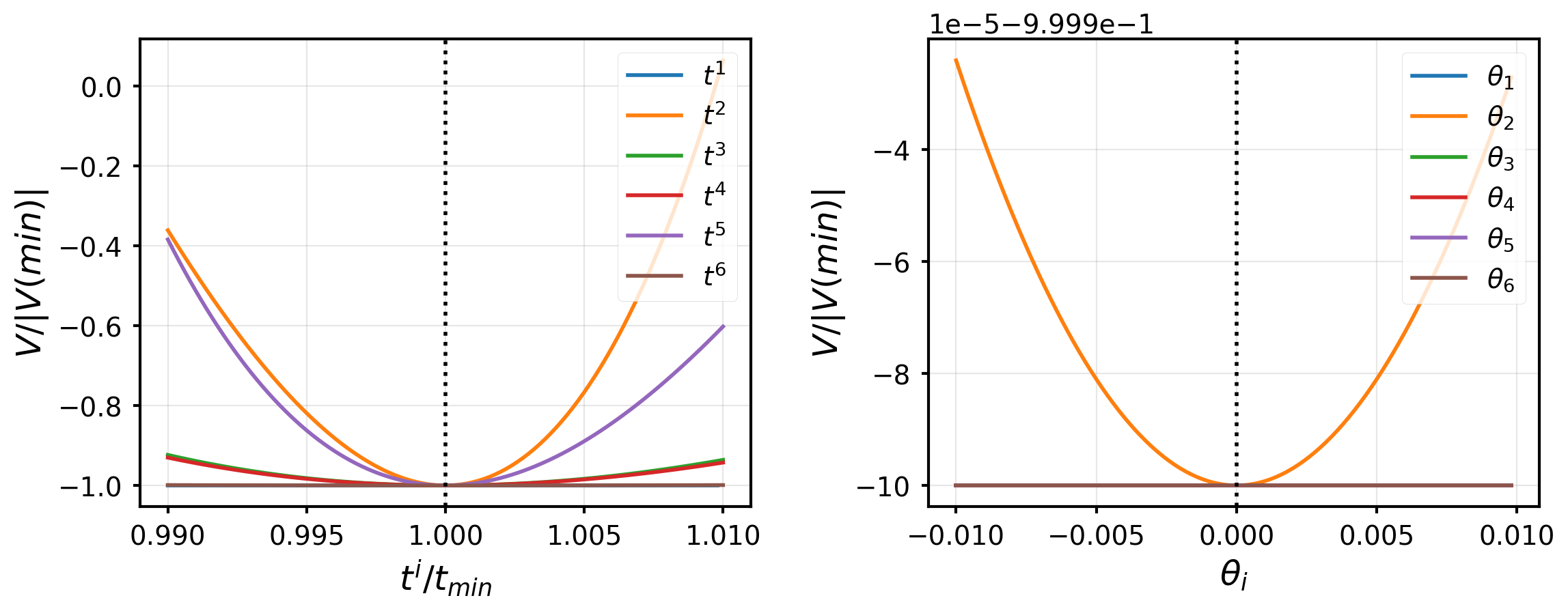}
    \caption{LVS minimum identified in a $h^{1,1}=6$ scenario through direct minimisation of the scalar potential specified by the parameters in Eq.~\eqref{eq:HybridExPar}.}\label{fig:LVS-example-large-h11}
\end{figure}

Our numerical framework readily identifies both pure LVS and these LVS-like hybrid solutions even at higher $h^{1,1}$. For example, we looked at the Calabi-Yau threefold with $h^{1,1}=6$ obtained from a triangulation of the polytope with vertices given by
\begin{equation}\label{eq:poly9543}
    \left(\;\begin{array}{cccccccccc}
    \phantom{-}0& \phantom{-}0&  -1 & \phantom{-}0 & \phantom{-}0 & \phantom{-}0 & \phantom{-}1 & \phantom{-}1 & -1 & \phantom{-}0\\
    \phantom{-}0& \phantom{-}0&  \phantom{-}0 & \phantom{-}0 & \phantom{-}1 & -1 & -1 & \phantom{-}0 & \phantom{-}0 & -1\\ 
    \phantom{-}1& -1& \phantom{-}0 & \phantom{-}0 & \phantom{-}0 & -1 & \phantom{-}0 & \phantom{-}0 & \phantom{-}0 & -1 \\  
    \phantom{-}0& -1& \phantom{-}1 & \phantom{-}1 & \phantom{-}0 & \phantom{-}0 & \phantom{-}0 & \phantom{-}0 & \phantom{-}0 & -1 \\
    \end{array}\;\right)\,.
\end{equation}
For this geometry, we obtained the solution shown in Fig.~\ref{fig:LVS-example-large-h11} for the parameter choice
\begin{equation}\label{eq:HybridExPar}
    g_s=0.03\, ,\; W_0 = -10^{0.5}\, ,\; A_i=1\, ,\; a_i=\dfrac{2\pi}{22}\, ,\;\alpha=0\, ,\;\chi(X)=-90\, .
\end{equation}
The minimum occurs at
\begin{equation}
    \mathcal{V}^{(0)} \approx 3{,}590
\end{equation}
and vacuum energy
\begin{equation}
    \langle V\rangle \approx -3.0\times10^{-11} \, .
\end{equation}
The blow-up modulus $\tau_s= \tau_2$ satisfies
\begin{equation}
    \tau_s \approx 21.4 \approx 0.7 a_s^{-1}\ln(\mathcal{V})\, ,
\end{equation}
while other divisor volumes $\tau_{i\neq s} \gtrsim 150$ are much larger.

For this geometry, we obtained the solution shown in Fig.~\ref{fig:LVS-example-large-h11}. Here, $\tau_s=\tau_2$ stabilises at hierarchically smaller values than the other five moduli, characteristic of LVS stabilisation where the $\tau_s$ non-perturbative term competes with leading-order contributions.

\subsection{Coexisting Vacua in the Landscape}
\label{subsec:double-mins}

Our numerical searches revealed the intriguing phenomenon of multiple minima coexisting within the same scalar potential for specific $(g_s, W_0)$ parameter values.\footnote{This feature, exhibiting multiple solutions for a single set of parameters has also been observed in~\cite{AbdusSalam:2007pm}.}
We identified several distinct coexistence scenarios; LVS minima paired with KKLT solutions, and LVS minima coexisting with K\"ahler uplifted solutions. 
Notably absent were potentials containing both KKLT and K\"ahler uplifted minima simultaneously; a finding consistent with the theoretical framework developed in \S\ref{subsubsec:Westphal-derv} and the numerical results presented in \S\ref{subsec:known-solns-results}. These multiple-minima configurations appear generically across various manifolds, requiring only the existence of LVS-type solutions.

\begin{figure}[t!]
    \centering
    \includegraphics[scale=0.4]{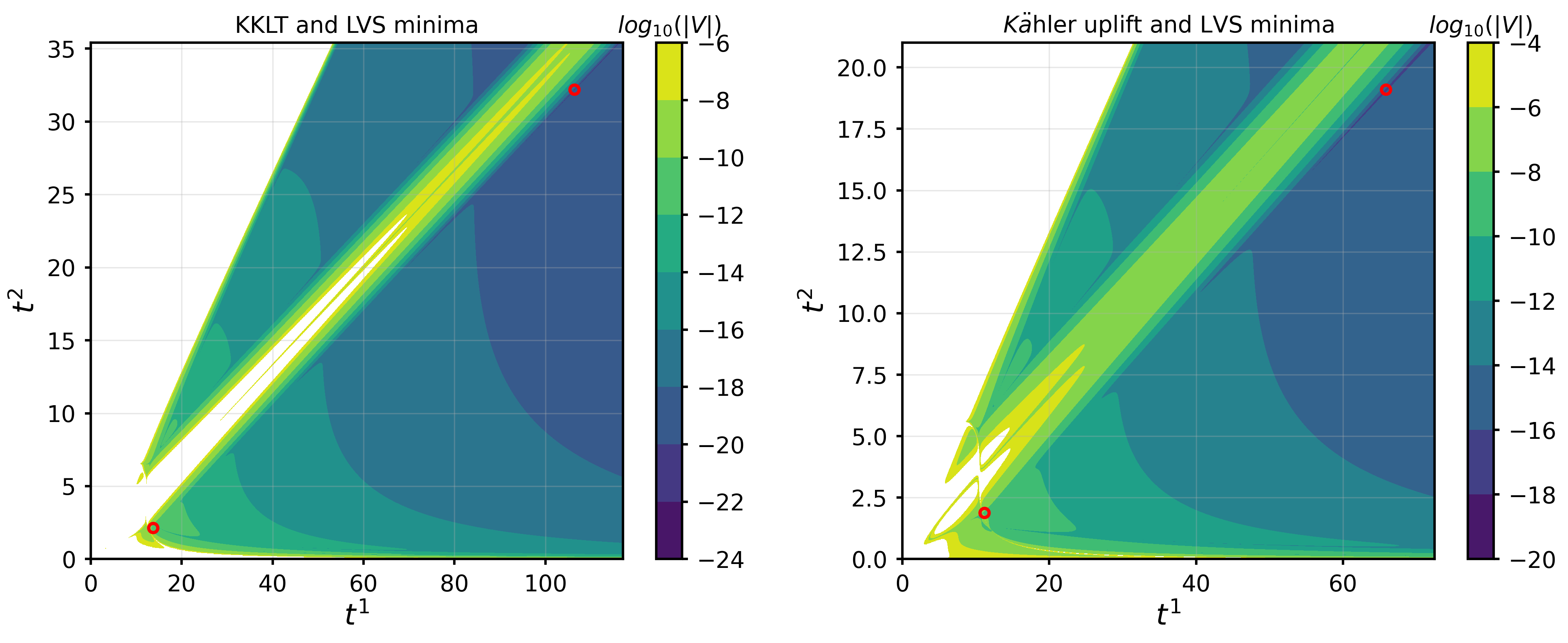}
    \caption{Coexisting minima in the scalar potential for $\mathbb{P}^4_{[1,1,1,6,9]}[18]$ at $h^{1,1}=2$ with K\"ahler cone condition $0<t_2<\frac{1}{3}t_1$ and with parameters $A_i=\{1.11,1\}$, $a_i=\{2\pi/24,2\pi/22\}$, $\chi=-540$, $\mathrm{e}^{K_{\text{cs}}}=0.03$, and $\alpha=10^{-6}$. \emph{Left:} Potential with $g_s=0.055$, $W_0=-0.024$ exhibiting both a SUSY KKLT minimum at $\mathcal{V}^{(0)}=112$ ($\langle V\rangle =-3.46\times10^{-11}$) and an LVS minimum at $\mathcal{V}^{(0)}=67,\!000$ ($\langle V\rangle =-4.77\times10^{-21}$). \emph{Right:} Potential with $g_s=0.075$, $W_0=-0.192$ containing a K\"ahler uplifted de Sitter minimum at $\mathcal{V}^{(0)}=68$ ($\langle V\rangle =+4.22\times10^{-9}$) alongside an LVS minimum at $\mathcal{V}^{(0)}=16,\!000$ ($\langle V\rangle =-2.92\times10^{-17}$).}
    \label{fig:doublemin-example}
\end{figure}

Our identification protocol builds upon the minimisation and filtering procedures from \S\ref{subsec:known-solns-results}. We systematically compare solutions across three categorised lists (KKLT, LVS, and K\"ahler uplifted) for the same values of $(g_s, W_0)$.
Solutions consistently appear in either KKLT-LVS or K\"ahler uplifted-LVS pairs, with the LVS solution always occupying the larger volume region. The coexistence of an LVS solution with a partner minimum depends primarily on the value of $|W_0|$. Indeed, for KKLT-LVS pairs, the KKLT solution exists when $|W_0|$ is sufficiently small to produce volumes $\mathcal{V}^{(0)}>\delta \mathcal{V}(t^i,g_s)$. When present, these KKLT solutions naturally coexist with LVS minima in the same potential. In contrast, for K\"ahler uplifted-LVS pairs, K\"ahler uplifted solutions typically exceed the KKLT threshold on $|W_0|$ and hence they can pair with LVS minima at larger $|W_0|$ values. The detailed evolution of solution types with varying $|W_0|$ is analysed in \S\ref{subsec:vary-W0-results}.

LVS solutions consistently appear at larger volumes, but the relative depths of minima vary. In KKLT–LVS pairs (both AdS), the KKLT minimum is always deeper and at smaller volume. In contrast, Kähler uplifted–LVS pairs show more varied behaviour due to the uplift mechanism, ranging from deep AdS to Minkowski, or dS minima, depending on the uplift strength.\footnote{In some cases, full dS uplift is not achieved before a runaway direction develops, leading to higher AdS minima compared to the LVS solution instead.}

This progression yields several possible double-minima configurations within the same potential such as: two AdS minima with the smaller-volume solution deeper (KKLT-LVS or K\"ahler Uplifted-LVS pairs),  two AdS minima with the larger-volume solution deeper (K\"ahler Uplifted-LVS pairs), Minkowski (K\"ahler Uplifted) + AdS (LVS) pair, and dS (K\"ahler Uplifted) + AdS (LVS) pair. An example of both a KKLT-LVs (AdS - AdS) pair and a K\"ahler Uplifted - LVS (dS - AdS) pair are shown in Fig.~\ref{fig:doublemin-example}.

\begin{figure}[t!]
    \centering
    \includegraphics[scale=0.55]{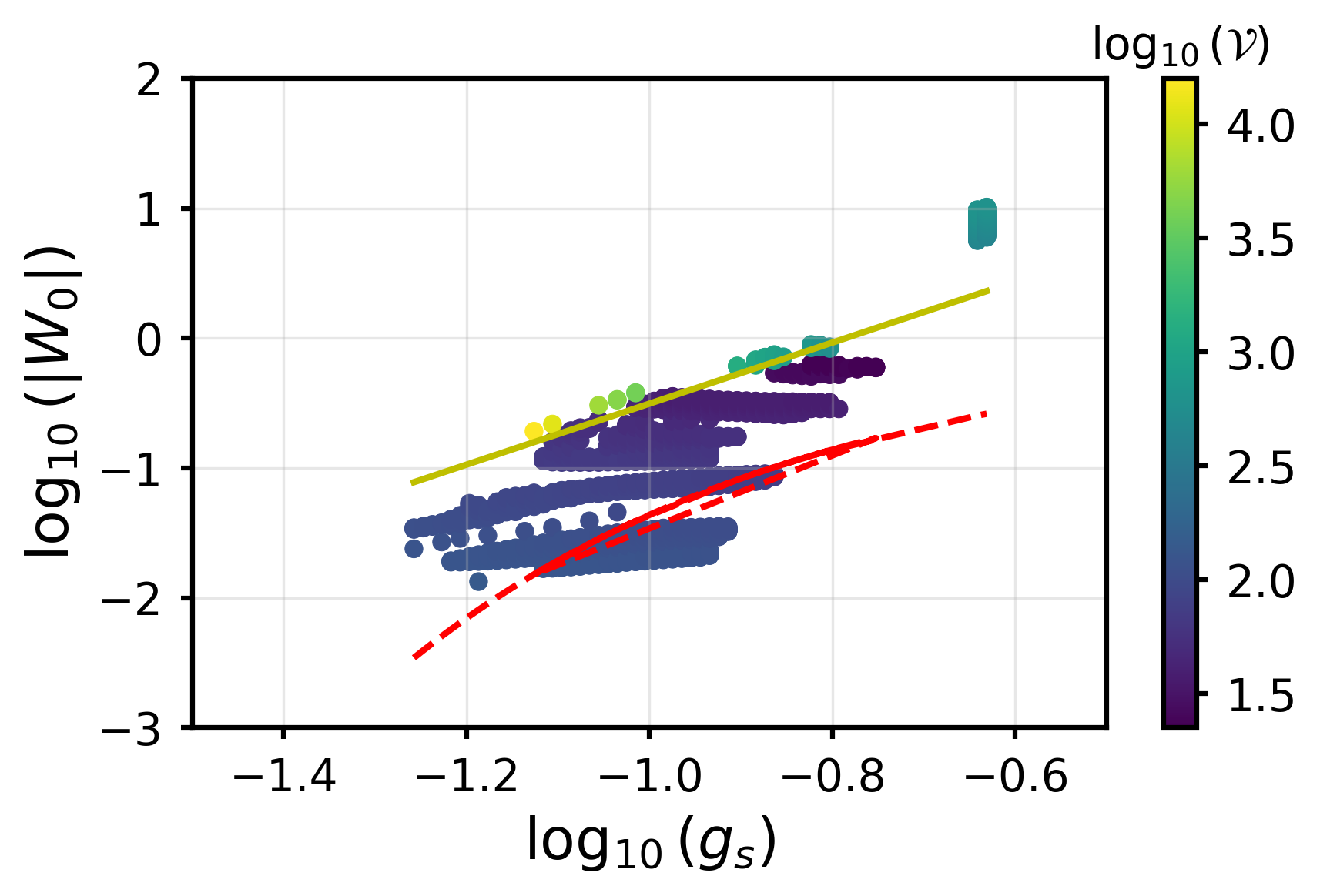}
    \caption{Parameter space $(g_s, W_0)$ exhibiting coexisting minima in compactifications on $\mathbb{P}^4_{[1,1,1,6,9]}[18]$. Colours indicate the volume of the LVS solution in each double-minimum configuration. The evolution of LVS volumes with varying $|W_0|$ is analysed in detail in \S\ref{subsec:vary-W0-results}. Parameters: $h^{1,1}=2$, $A_i=\{1.11,1\}$, $a_i=\{2\pi/24,2\pi/22\}$, $\chi=-540$, $\mathrm{e}^{K_{\text{cs}}}=0.03$, and $\alpha=10^{-6}$. 
    }
    \label{fig:doublemin-scan}
\end{figure}

Fig.~\ref{fig:doublemin-scan} displays the regions in the $(g_s, W_0)$ parameter space where double minima were identified for the $\mathbb{P}^4_{[1,1,1,6,9]}[18]$ Calabi-Yau manifold. In this scan, we sampled 200 values of $g_s$ in the range $(0.03,0.5)$, and for each, 5,000 values of $-|W_0|$ in the range $(10^{-3}, 10^{2})$, with all other parameters fixed as specified in the caption of Fig.~\ref{fig:doublemin-scan}.
This produced approximately $4.5 \times 10^5$, $2.5 \times 10^5$, and $6.6 \times 10^5$ solutions. After removing SUSY KKLT solutions, we found 14,819 instances of multiple minima arising within the same potential.

\subsection{Adding explicit uplifting sources}
\label{subsec:h11-3+_and_uplift}

Thus far, we have focused exclusively on the $F$-term potential. 
In typical compactifications, additional uplifting terms, e.g. from anti-D3 branes, T-branes or D-terms, arise that can raise AdS minima to de Sitter. 
These uplift contributions scale as $\sim\mathcal{V}^{-n}$, $n<3$, with the overall volume $\mathcal{V}$.

Indeed, the introduction of an explicit uplifting term in our analysis enables new possibilities.
Specifically, we will consider anti-D3 brane uplifts\footnote{An explicit bound on the net D3-tadpole required to maintain control over warping corrections in LVS in the presence of anti-D3 branes has been derived in \cite{Gao:2022fdi}. Similarly, control issues in KKLT due to warping effects have been identified e.g. in the context of the singular bulk problem \cite{Carta:2019rhx,Gao:2020xqh,Carta:2021lqg}, see also \cite{Moritz:2025bsi} for a strengthened version of the problem.}
which lead to an additional contribution
\begin{equation}\label{eq:uplift-term}
V_{\text{up}} = \frac{D_{\text{up}}}{\mathcal{V}^{4/3}} > 0
\end{equation}
in the full scalar potential. We keep the coefficient $D_{\text{up}}$ here as a free parameter, and simply note that it can in principle be computed in explicit examples where warped Klebanov-Strassler throats \cite{Klebanov:2000hb,Giddings:2001yu} can be realised, see in particular \cite{McAllister:2024lnt} and \cite{Crino:2020qwk} for recent discussions in the context of KKLT and LVS respectively.

\begin{figure}[t!]
\centering
\includegraphics[scale=0.42]{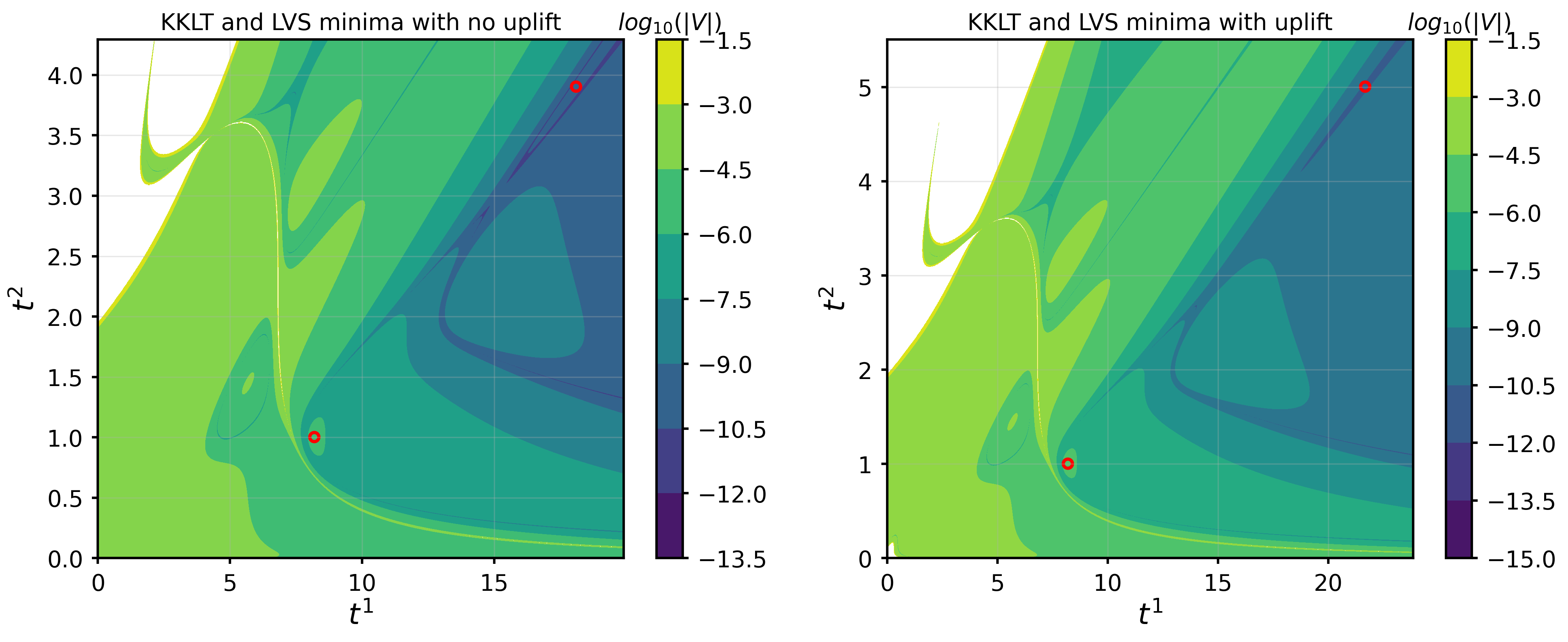}
\caption{Uplifted KKLT-LVS double minimum configuration. Starting from an AdS pair ($D_{\text{up}}=0$) with KKLT minimum at $\mathcal{V}=22.796$ ($V=-1.1971\times10^{-6}$) and LVS minimum at $\mathcal{V}=315$ ($V=-8.66\times10^{-11}$), we introduce the uplifting term (Eq.~\eqref{eq:uplift-term}) and increase $D_{\text{up}}$ until achieving a dS LVS solution. At $D_{\text{up}}=2.82\times10^{-7}$, the LVS minimum uplifts to $\mathcal{V}=549$ ($V=+1.94\times10^{-11}$) while the KKLT minimum remains nearly unchanged at $\mathcal{V}=22.802$ ($V=-1.1928\times10^{-6}$). Computed for $\mathbb{P}^4_{[1,1,1,6,9]}[18]$ with $g_s=0.18$, $W_0=-0.6$, and other parameters matching Fig.~\ref{fig:doublemin-scan}.}
    \label{fig:KKLT-LVS-double-soln-uplift}
\end{figure}

Adding an explicit uplift term unlocks vacuum pairings beyond those found in \S\ref{subsec:double-mins}. In particular, we can now engineer a small-volume AdS minimum coexisting with a larger-volume Minkowski or de Sitter vacuum --- configurations unattainable by Kähler uplifting alone. As described in \S\ref{subsubsec:KKLT-results} and \S\ref{subsubsec:LVS-results}, LVS and KKLT minima often coexist for the same $(g_s, W_0)$. After applying the uplift of Eq.~\eqref{eq:uplift-term}, the LVS branch becomes a metastable dS solution while the KKLT branch remains a deep, slightly SUSY-broken AdS vacuum. In \S\ref{subsec:tunnelling} we analyse the resulting tunnelling dynamics. More generally, by varying the uplift strength one can realise any combination of AdS, Minkowski, and dS minima at both small and large volumes in a single potential.

\begin{figure}[t!]
    \centering
    \includegraphics[width=0.48\linewidth]{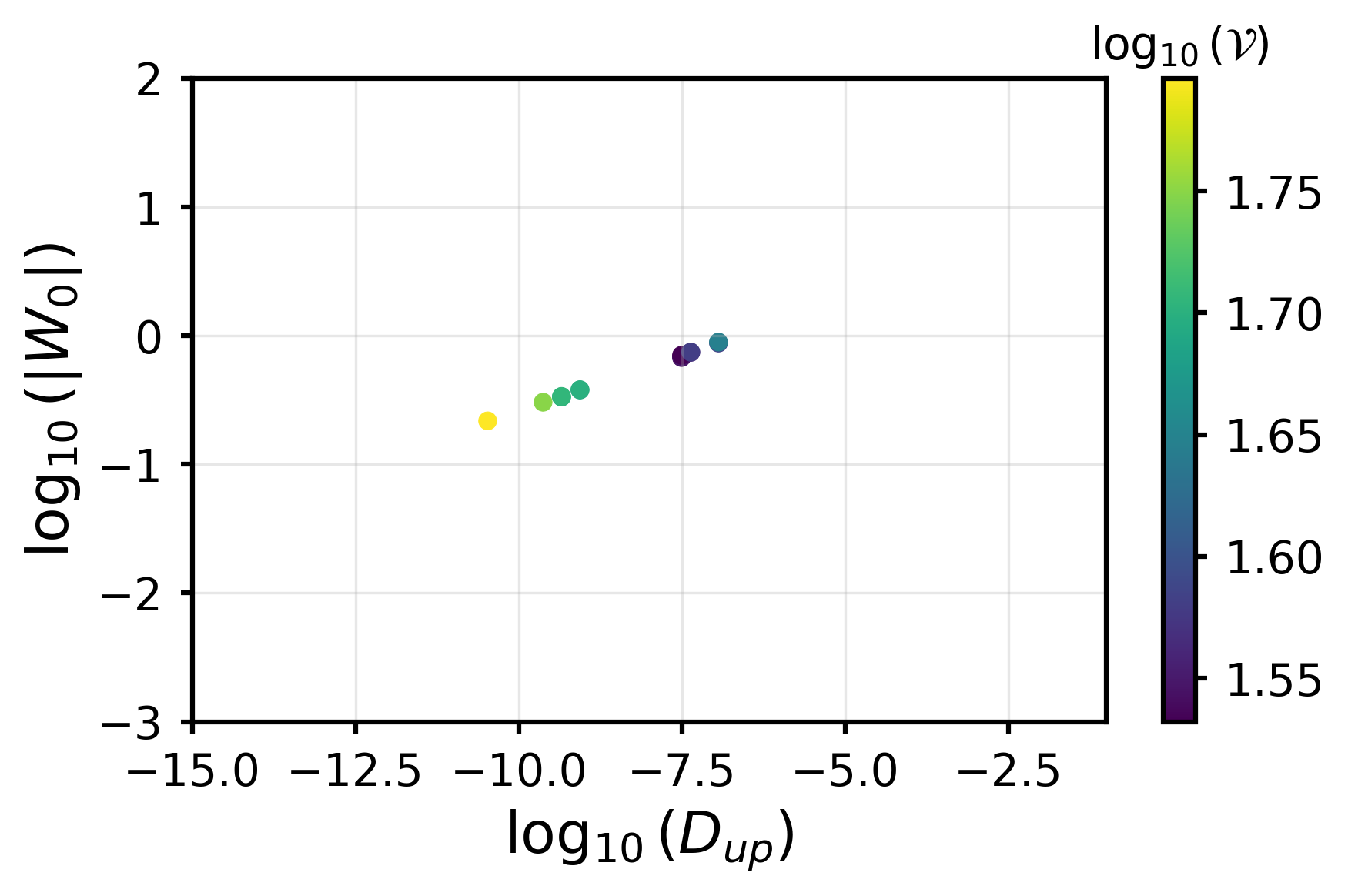}
    \caption{Parameter combinations yielding multiple minima for compactifications on $\mathbb{P}^4_{[1,1,1,6,9]}[18]$ that can be simultaneously uplifted to dS via Eq.~\eqref{eq:uplift-term}. The discrete scattering of values reflect the finite resolution of our numerical scan; the underlying behaviour is expected to be continuous. Each point corresponds to a point in the parameter space where double minima are found, the colour corresponds to the volume of the smaller solution.
    }
    \label{fig:West_LVS_uplift_scan}
\end{figure}

We explore the landscape of double dS solutions by first considering the pairs of double minima we found in Fig.\ref{fig:doublemin-scan} (where we in effect had $D_{\text{up}}=0$). This provides a start point for us to run two simultaneous `iterative' optimisation processes that takes the non-uplifted minimum and tracks it as we slowly increase $D_{\text{up}}$, in a procedure reminiscent of how we found the K\"ahler Uplifted solutions by increasing $|W_0|$ when starting from a KKLT solution. The algorithm continues to increase the uplift parameter until one of our solutions disappears (in effect the potential is now runaway at that point). The solutions shown in Fig.~\ref{fig:West_LVS_uplift_scan} correspond to the double minima found in Fig.~\ref{fig:doublemin-scan} that survived uplifting both solutions to dS. All pairs of solutions that could be uplifted to double dS minima contained a combination of K\"ahler uplifted and LVS-type solutions; all KKLT minima were too deep and the uplift required to bring them to dS caused the LVS solution to be uplifted to a runaway. Shown in Fig.~\ref{fig:KKLT-LVS-double-soln-uplift} is an example of a KKLT-LVS solution pair, we can seen in the second plot that when $D_{\text{up}}\sim 10^{-7}$ we have sufficient uplift to promote the LVS solution to dS, but the KKLT solution is only effected at the 4th significant figure.

\section{Dynamics in the string landscape}
\label{sec:P11169}

This section presents applications of our numerical framework that showcase its utility in studying the string theory landscape. Our approach enables both efficient large-scale parameter scans and detailed investigations of vacuum dynamics and stability. The framework is highly parallelisable, allowing for extensive scans over values of the flux-induced superpotential $W_0$. This facilitates systematic studies of vacuum properties across wide regions of parameter space, and helps identifying regions that support de Sitter solutions or exhibit interesting phenomenological features. Building on these results, our methods provide a foundation for studying dynamical processes in the landscape --- such as vacuum transitions and tunnelling.

\subsection{Vacuum structure from $|W_0|$ variation}
\label{subsec:vary-W0-results}

We now proceed to examine the transitions between different vacuum types by systematically scanning the flux parameter $W_0$. In KKLT solutions, the overall volume scales exponentially with the divisor volumes $|W_0| \sim \mathrm{e}^{-a_n \tau_n}$, recall Eq.~\eqref{eq:trivial-moduli-relation-ii}, whereas in the LVS regime one finds the opposite scaling behaviour $ |W_0| \sim \mathcal{V}$ (recall Eq.~\eqref{eq:LVS_solution}).
Our numerical framework allows us to probe these competing scalings within a single scalar potential by tracking, for example, how the LVS minimum deepens and then gives way to a KKLT minimum as $|W_0|$ is decreased. This allows us to understand the continuous deformation between the two solution branches and the parameter windows in which they coexist.

\begin{figure}[t!]
    \centering
    \includegraphics[width=\linewidth]{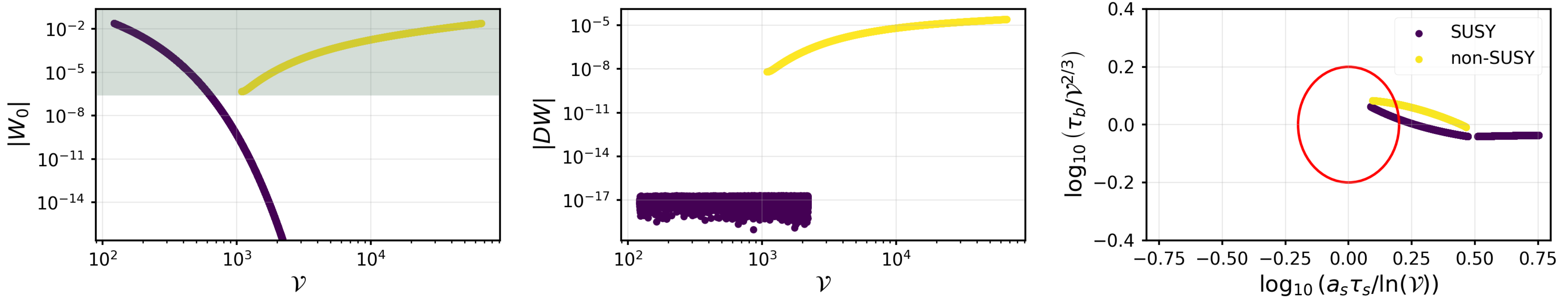}
    \caption{Evolution of both KKLT and LVS solution with decreasing $|W_0|$. Starting from an initial value $|W_0|\sim 10^{-2}$ where both solutions coexist, we observe opposite scaling as we dial the flux superpotential down to $|W_0|\sim 10^{-15}$.
    \emph{Left:} Beginning from either the KKLT branch (top left) or the LVS branch (top right), lowering $|W_0|$ causes the KKLT volume to grow, while the LVS volume shrinks. The shaded region shows the values of $|W_0|$ for which both vacua coexist. 
    \emph{Middle:} The scale of SUSY breaking $|DW|=\sum_i |D_iW|$ as a function of the volume.
    \emph{Right:} Deviation from characteristic LVS scaling $\tau_b\sim\mathcal{V}^{2/3}$, $a_s\tau_s\sim\ln\mathcal{V}$. 
    }
    \label{fig:LVS_soln_lowering_W0}
\end{figure}

Let us first test the prediction that an LVS AdS vacuum shrinks as $|W_0|$ decreases, whereas below some critical $|W_0|$ value a KKLT-like solution exists.
We start from a potential with coexisting LVS and supersymmetric KKLT minima obtained for flux compactifications on $\mathbb{P}^4_{[1,1,1,6,9]}[18]$ using the same parameters as in Fig.~\ref{fig:doublemin-example}, with $\tau_1\equiv\tau_b$ and $\tau_2\equiv\tau_s$.
We then tracked the volume of each solution separately as $| W_0|$ was lowered.
As shown in Fig.~\ref{fig:LVS_soln_lowering_W0}, the volume indeed decreases when starting from the LVS solution (top right), while the KKLT volume increases (top left).
Interestingly, the LVS solution branch eventually disappears below some threshold value $|W_0|\sim 10^{-7}$ at which point the KKLT branch takes over.

\begin{figure}[t!]
    \centering
    \includegraphics[width = 0.6\linewidth]{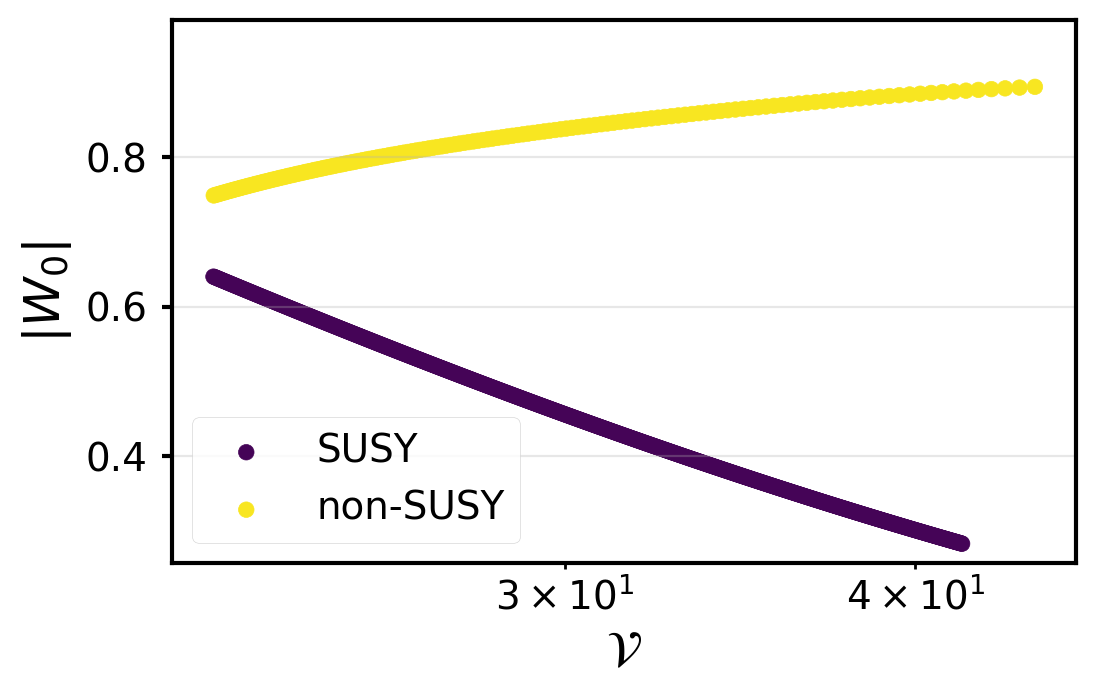}
    \caption{Evolution of a small-volume de Sitter vacuum, coexisting with an LVS minimum, as $|W_0|$ is decreased. The solutions are computed using the iterative procedure of \S\ref{subsec:NT-reducing_computations_per_run} for compactifications on $\mathbb{P}^4_{[1,1,1,6,9]}[18]$ with the same parameters as in Fig.~\ref{fig:LVS_soln_lowering_W0}. Note that the LVS minimum remains present across the entire scanned range of $|W_0|$.
    }
    \label{fig:west_soln_lowering_W0}
\end{figure}

We can similarly investigate the reverse transition, with increasing $|W_0|$, the vacuum shifts from a KKLT AdS minimum to a non-supersymmetric Kähler-uplifted solution. As anticipated in \S\ref{subsubsec:Westphal-derv}, once $|W_0|$ exceeds a critical value, the family of KKLT solutions ceases to exist and is replaced by a SUSY-breaking minimum.  Fig.~\ref{fig:west_soln_lowering_W0} illustrates this behaviour: starting from a small-volume dS vacuum at large $|W_0|$, lowering $|W_0|$ suppresses the uplift, first driving the solution to AdS, then causing the Kähler-uplift branch to disappear below which the familiar KKLT AdS vacuum reemerges, satisfying $D_iW=0$ and exhibiting the expected scaling.

Interestingly, the transition occurs while LVS solutions persist in the potential, enabling observation of various coexisting minima configurations through $|W_0|$ variation alone:\footnote{Recall that what we call small-volume is still larger than the string scale.}
\begin{itemize}
    \item dS (small-volume) + AdS (LVS) pairs
    \item Minkowski (small-volume) + AdS (LVS) pairs
    \item AdS (small-volume deeper) + AdS (LVS) pairs
    \item AdS (KKLT deeper) + AdS (LVS) pairs
\end{itemize}
Once uplifting terms are added, all combinations of dS/Minkowski/AdS exist.

\subsection{Tunnelling and Stability}
\label{subsec:tunnelling}

A second application of our results concerns vacuum decays. The stability of vacua in the string landscape, including those identified in our analysis, has been the subject of sustained investigation \cite{Kachru:2003aw,Westphal:2005yz,deAlwis:2013gka,Freivogel:2005vv,Marsh:2015zoa,Johnson:2008vn,Aguirre:2009tp,Brown:2011ry}. Previous studies have examined both KKLT and LVS minima in a range of configurations, considering their uplifted de Sitter versions (analysing tunnelling to larger-volume solutions or decompactification) as well as their non-uplifted anti-de Sitter counterparts.

To date, the discussion has primarily focused on two decay channels:
\begin{enumerate}
\item decay of a candidate de Sitter vacuum towards the runaway decompactification regime, studied via the Coleman–De Luccia formalism \cite{Coleman:1980aw}, and
\item transitions between distinct flux configurations, analysed using the Brown-Teitelboim formalism \cite{Brown:1988kg} of bubble nucleation, which determines viable decay channels and corresponding vacuum lifetimes.
\end{enumerate}
The decay probability per unit volume and time follows the standard exponential form
\begin{equation}\label{eq:BT-prob}
	\Gamma \sim \mathrm{e}^{-B}, \quad B = S_{\text{instanton}} - S_{\text{background}}\, ,
\end{equation}
where $S$ denotes the Euclidean action.
In general, decay towards decompactification, although exponentially suppressed, tends to dominate over decay to other flux vacua.\footnote{In \cite{Johnson:2008vn,Aguirre:2009tp}, it was argued that an instanton connecting two finite-volume vacua may not exist, potentially obstructing the population of the landscape. See also the subsequent analysis in \cite{Brown:2011ry}.}

Here, we propose a third decay channel that has not been previously considered. In Sections~\S\ref{subsec:double-mins} and~\S\ref{subsec:h11-3+_and_uplift}, we demonstrated the existence of configurations with multiple minima arising from fixed flux choices, in which an LVS-type minimum coexists with either a KKLT or Kähler uplifted one.

The availability of explicit examples of scalar potentials in which all saddle points are known and which, in principle, span different energy scales, offers a rich theoretical laboratory for investigating the population of the landscape and cosmological implications. A comprehensive analysis of decay rates between these vacua and their broader physical consequences lies beyond the scope of this work. Here, we restrict ourselves to outlining some of their key potential implications:
\begin{itemize}
\item {\bf Vacuum decay towards another local minimum.} Concrete calculations can be performed in order to determine the transition rate from a high energy de Sitter minimum to the second local minimum for which the cosmological constant can be naturally tuned by the fluxes to have zero or very small value. The decay rate to decompactification can be estimated in a similar way. Notice that we have found all possible combinations in which the higher minimum corresponds to smaller or larger volume. Naively the decay rate from a de Sitter vacuum to a Minkowski vacuum is given by
\begin{equation}
    \Gamma=\exp\biggl (-\frac{S(\varphi_0)}{1+\left(\tfrac{4V_0}{3\sigma^2}\right)^2}\biggl )\, , \;\qquad \sigma=\int_{\varphi_0}^{\varphi_1}d\varphi \sqrt{2V(\varphi)}\, ,
\end{equation}
where $S(\varphi_0)=24\pi^2/V_0$ and $\sigma$ is the bubble's tension with $\varphi_1$ the location of the second local minimum. In order to compare this decay rate to the one towards the runaway minimum, we would need to compare the values of the wall tension in both cases, the value of $V(\varphi_0)$, the value of the potential at the local maximum $V(\varphi_{max})$ or saddle that separates the two local minima in order to estimate the height of the barrier, the validity of the thin-wall approximations, etc. In particular, we would need to compare the values of $\Delta\varphi \sqrt{V_{max}}$ for both transitions in order to compare the wall's tension in both cases. Hawking-Moss transitions may also be implemented. All of this varies from case to case depending on the particular compactification and flux choice and requires a systematic analysis. 
Furthermore, the coexistence of Minkowski and AdS minima allows us to explicitly provide a flux compactification realising the second example discussed in \cite{Coleman:1980aw}, decay from Minkowski towards an AdS crunch  for which $\Gamma$ is as above by changing the plus by a minus sign in the denominator.
\item {\bf Runaway dilatonic domain walls.} A quantitative analysis can be done for the existence or not  of the corresponding bounce solution connecting the two local minima in the presence of the runaway, extending  the discussion on runaway dilatonic domain walls in \cite{Johnson:2008vn,Aguirre:2009tp}.

\item {\bf Not bubble of nothing decay.}
The fact that the value of the potential at the LVS minimum decreases with decreasing $W_0$ would naively lead one to believe that the minimum will continue decreasing until it disappears to an unbounded from below potential. This is what happens in 6d and it has been given an interpretation to a decay towards a bubble of nothing \cite{Brown:2011gt}. However, here we confirm the argument of  \cite{deAlwis:2013gka} in which, at a critical value of $W_0$, the LVS minimum disappears but a KKLT minimum appears and as $W_0$ decreases further the value of the potential at the minimum increases rather than decreases, avoiding the bubble of nothing decay.

\item {\bf Potential instability of AdS minima.}
For the case of double AdS minima, where either K\"ahler uplifted-LVS or SUSY KKLT-LVS pairs coexist within the same potential, Coleman-de Luccia transitions from LVS vacua to deeper minima at smaller volumes may occur, establishing the LVS solution as metastable. Notice that in this case there is no transition towards decompactification (following the standard Coleman-de-Luccia prescription, see however \cite{Cespedes:2023jdk}). Since there is a dominant decay mode to another AdS this may have impact in attempts to construct CFT duals of these AdS LVS minima \cite{Conlon:2018vov, deAlwis:2014wia}.

\item {\bf Volume modulus inflation.}
The tuning of the fluxes that gives rise to two local (non-runaway) minima can be used to smoothly deform the higher minimum such that this minimum disappears and gives rise to a potential which is naturally flat at high energies providing an excellent candidate to realise volume modulus inflation \cite{Conlon:2008cj}. As usual in the landscape, the tuning is given by the richness of flux vacua. This may also lead to a period of kination and its many potential cosmological implications as recently studied in \cite{Conlon:2022pnx,Apers:2022cyl,Conlon:2024uob,Revello:2024gwa,Ghoshal:2025tlk}.

\item {\bf AdS wormholes and inflation.} 
The fact that we can also get coexisting dS and AdS vacua provides a concrete example of the scenario proposed in \cite{Betzios:2024oli,Betzios:2024iay} in which an Euclidean AdS wormhole can give rise to a natural period of inflation with the dS vacuum responsible for the current acceleration. This is an interesting alternative proposal for initial conditions as compared to the Vilenkin, Hartle-Hawking proposals for the wave function of the universe. Note also that in  our examples we have both cases with the dS corresponding either to the smaller or the larger volume case.

\end{itemize}

\section{Conclusions}
\label{sec:conclusions}
We have developed a unified computational framework for assembling and evaluating the full scalar potential in Type IIB flux compactifications, including both perturbative and non-perturbative ingredients. This implementation leverages just-in-time compilation, and automatic differentiation based on the JAX environment \cite{jax2018github} to deliver an efficient, scalable toolkit for systematic landscape explorations. As a proof of principle, we have used this implementation to carry out non-exhaustive searches for vacua across a range of Calabi-Yau orientifolds with $h^{1,1}\leq 6$.

Our large-scale numerical study of Type IIB flux compactifications has uncovered new features of vacuum structures which may have implications on the stability of such solutions in the string landscape. By employing the optimised framework described in \S\ref{sec:numerics}, we systematically surveyed moduli stabilisation scenarios on Calabi-Yau threefolds with $2 \le h^{1,1}\le6$. In particular, we confirm that as $|W_0|$ is dialled down from large values, the LVS AdS minimum deepens until it ceases to exist, at which point a KKLT-like minimum emerges and its energy increases as $|W_0|$ continues to decrease.

We further establish that distinct stabilisation mechanisms, specifically KKLT, LVS and Kähler uplifted, often coexist for the same choice of flux parameters (cf.~\S\ref{subsec:double-mins}), even in the presence of explicit uplifting contributions (cf.~\S\ref{subsec:h11-3+_and_uplift}). Across all examined Calabi-Yau geometries, pairs of minima with any combination of AdS, Minkowski, or dS vacua can arise simultaneously. The generic occurrence of multiple local vacua at a single point in parameter space opens new avenues for vacuum selection dynamics, tunnelling phenomena, and novel inflationary scenarios.

Our results reveal a richer web of interconnected vacua in the string landscape. The numerical toolkit introduced here offers a potent platform for systematically charting this complex terrain.

Looking ahead, there are several exciting avenues to extend this work. First, integrating our JAX-based Kähler-moduli implementation with complementary packages such as JAXVacua \cite{Dubey:2023dvu} would enable fully unified scans that stabilise the axio-dilaton, complex structure, and Kähler moduli in one pass, opening the door to systematic investigations of backreaction effects. Second, the modular design of our  framework means that additional corrections --- such as higher-order loop and $\alpha'$ effects, brane-induced threshold corrections, or alternative uplifts like T-branes --- can be included with minimal effort, allowing a quantitative assessment of control and convergence. Third, the discovery of coexisting minima suggests fertile ground for studying vacuum dynamics: by computing tunnelling actions between vacua in a fixed potential, one could explore new channels for vacuum decay and cosmological transitions within a single flux choice. Finally, pushing to larger $h^{1,1}$ and incorporating additional search strategies, maybe even guided by Machine Learning, may uncover novel corners of the landscape, bringing us closer to concrete, fully stabilised de Sitter constructions.

\section*{Acknowledgements}

We  would like to thank Michele Cicoli, Pellegrino Piantadosi, Mario Ramos-Hamud and Gonzalo Villa for useful conversations. CH thanks K. Önder and N. McStay for multiple helpful discussions related to this work. We thank Panos Betzios, Ioannis Gialamas and Olga Papadoulaki for communications about their work in \cite{Betzios:2024oli,Betzios:2024iay}. The work of CH has been partially supported by Cambridge Trust, NUI and by STFC consolidated grants ST/T000694/1 and ST/X000664/1. The work of FQ has been partially supported by STFC consolidated grants ST/P000681/1, ST/T000694/1 and by a NYUAD research grant. 

\appendix

\section{Explicit formulas for K\"ahler metric and the potential}\label{App:Kmet}

Starting with a K\"ahler potential of eq. \eqref{eq:Kpot-defs} and considering perturbative corrections $\delta \vol(\vol,s)$ as defined in Eq.~\eqref{eq:BBHL+log-pert-corrections}, we calculate analytic expression for the K\"ahler metric and the corresponding terms in the scalar potential. First, we find
\begin{equation}\label{eq:Kmet_down_appendix_def}
K_{T_n \bar{T}_m} = c_1 t^m t^n +c_2 \frac{\partial t^n}{\partial \tau_m}\, , \;\qquad K_{\tau \bar{\tau}}  = \frac{1}{4s^2} + c_{\tau}\, , \;\qquad K_{T_n\bar{\tau}} = c_{T} t^n\, ,
\end{equation}
where
\begin{align}
    c_1 &= \frac{1}{4\vol^2}\left( (1 + \delta \vol_\vol)^2 - \vol\, \delta \vol_{\vol\vol} \right)\, ,\;\qquad c_2 = -\frac{(1 + \delta \vol_\vol)}{4\vol}\, ,\\[0.3em]
    c_{\tau} &= \frac{\left(\delta \vol_s^2-\delta \vol_{ss} \vol \right)}{2\vol^2 }\, ,\;\qquad  c_{T} =\frac{\vol\,\delta \vol_{\vol s} -(1+\delta \vol_\vol ) \delta \vol_s }{4\vol^2} \, .
\end{align}
This can be checked against the calculations of \cite{AbdusSalam:2020ywo} where only the BBHL correction was considered.

Inverting the Kähler metric and using the expression for the $F$-terms in Eq.~\eqref{eq:FtermsExplicit}, we compute the coefficient functions $\phi_n$ in the expression for the scalar potential \eqref{eq:master-eqn} which are given by
\begin{equation}\label{eq:master-eqn_app_phi}
\begin{split}
\phi_0 
&= 
\frac{3\mathcal{V}^{(0)} (1 + \delta \vol_\vol)^2}{2\vol^2(6c_1\mathcal{V}^{(0)}+c_2)^2}\Biggl(c_2 +6\mathcal{V}^{(0)}(\mathcal{S}^{-1} c_{T}^2 + c_1)\Biggr) 
- 
3
\, ,\\
\phi_1^{(i)} 
&= 
\frac{a_i \tau_i(1 + \delta \vol_\vol)}{\vol(6c_1\mathcal{V}^{(0)}+c_2)^2}\Biggl(c_2 +6\mathcal{V}^{(0)}(\mathcal{S}^{-1} c_{T}^2 + c_1)\Biggr)\, ,
\\
\phi_2^{(i,j)} 
&=
\frac{4}{6c_1 \mathcal{V}^{(0)} + c_2}
\left[\frac{c_{T}^2}{S(6c_1\mathcal{V}^{(0)}+c_2)}  + \frac{c_1}{c_2}\right] \tau_i \tau_j 
+ 
 \frac{1}{c_2}\kappa_{ijk}t^k\, ,
\end{split}
\end{equation}
where we introduced, $\mathcal{S}$, is the Schur complement of the Kähler metric. 
\begin{equation}
\mathcal{S}= \frac{1}{4s^2} + c_{\tau} - \frac{6 \mathcal{V}^{(0)} c_{T}^2}{c_2(6\mathcal{V}^{(0)} c_1+c_2)}\, .
\end{equation}

\bibliographystyle{utphys}
\bibliography{biblio}

\end{document}